\documentclass[fleqn,usenatbib]{mnras}

\usepackage{newtxtext,newtxmath}
\usepackage[T1]{fontenc}

\DeclareRobustCommand{\VAN}[3]{#2}
\let\VANthebibliography\thebibliography
\def\thebibliography{\DeclareRobustCommand{\VAN}[3]{##3}\VANthebibliography}

\usepackage{graphicx}	% Including figure files
\usepackage{amsmath}	% Advanced maths commands
\usepackage{orcidlink}
\usepackage[para,flushleft]{threeparttable}
\usepackage{booktabs}
\usepackage{makecell}
\usepackage{multirow}
\usepackage{caption}
\usepackage{gensymb}
\usepackage{subcaption}
\usepackage{paralist}
\usepackage{lineno}
\title[K. Krishna et al.]{Insights from GRBs for optical follow-up of gravitational wave counterparts}

\author[K. Krishna et al.]{Kruthi Krishna\orcidlink{0000-0002-6967-5140},$^{1,2,3, 4}$\thanks{E-mail: kruthi.krishna@desy.de}
Andrew Levan\orcidlink{0000-0001-7821-9369},$^{4,5}$
Samaya Nissanke\orcidlink{0000-0001-6573-7773},$^{1,2,3,6,7}$
Morgan Fraser\orcidlink{0000-0003-2191-1674},$^{8}$
\newauthor 
Tomas Ahumada\orcidlink{0000-0002-2184-6430},$^{9}$ 
Shreya Anand \orcidlink{0000-0003-3768-7515}\thanks{LSST-DA Catalyst Postdoctoral Fellow},$^{10,11}$
Igor Andreoni\orcidlink{0000-0002-8977-1498},$^{12}$
Andreja Gomboc\orcidlink{0000-0002-0908-914X},$^{13}$
Mansi Kasliwal\orcidlink{0000-0003-3769-9559},$^{14}$
\newauthor
Andrea Melandri\orcidlink{0000-0002-2810-2143},$^{15}$
Silvia Piranomonte\orcidlink{0000-0002-8875-5453}$^{15}$ and
Patricia Schmidt\orcidlink{0000-0003-1542-1791},$^{16}$\\
$^{1}$Deutsches Elektronen-Synchrotron DESY, Platanenallee 6, D-15738 Zeuthen, Germany\\
$^{2}$Deutsches Zentrum für Astrophysik DZA, Postplatz 1, 02826 Görlitz, Germany\\
$^{3}$Institut für Physik und Astronomie, Universität Potsdam, 14476 Potsdam, Germany\\
$^{4}$Department of Astrophysics/IMAPP, Radboud University, P.O. Box 9010, 6500 GL, Nijmegen, The Netherlands\\
$^{5}$Department of Physics, University of Warwick, Coventry, CV4 7AL, UK\\
$^{6}$GRAPPA Institute, Anton Pannekoek Institute for Astronomy and Institute of High-Energy Physics,\\
University of Amsterdam, Science Park 904, 1098 XH Amsterdam, The Netherlands\\
$^{7}$Nikhef, Science Park 105, 1098 XG Amsterdam, The Netherlands \\
$^{8}$ School of Physics, University College Dublin, Belfield, Dublin 4, Ireland\\
$^{9}$Cerro Tololo Inter-American Observatory/NSF NOIRLab, Casilla 603, La Serena, Chile\\
$^{10}$Kavli Institute for Particle Astrophysics and Cosmology, Stanford University, 452 Lomita Mall, Stanford, CA 94305, USA\\
$^{11}$Department of Physics, Stanford University, 382 Via Pueblo Mall, Stanford, CA 94305, USA\\
$^{12}$Department of Physics and Astronomy, University of North Carolina at Chapel Hill, Chapel Hill, NC 27599-3255, USA\\
$^{13}$Center for Astrophysics and Cosmology, University of Nova Gorica, Vipavska 13, 5000 Nova Gorica, Slovenia\\
$^{14}$Division of Physics, Mathematics and Astronomy, California Institute of Technology, Pasadena, CA 91125, USA\\
$^{15}$ INAF – Osservatorio Astronomico di Roma, Via di Frascati 33, I-00078 Monteporzio Catone, Italy\\
$^{16}$School of Physics and Astronomy and Institute for Gravitational Wave Astronomy, University of Birmingham, Edgbaston, Birmingham, B15 2TT, UK
}
\date{Accepted XXX. Received YYY; in original form ZZZ}

\pubyear{2026}

\begin{document}
\label{firstpage}
\pagerange{\pageref{firstpage}--\pageref{lastpage}}
%\linenumbers
\maketitle

\begin{abstract}
% The abstract should briefly describe the aims, methods, and main results of the paper. It should be a single paragraph not more than 250 words (200 words for Letters). No references should appear in the abstract.

Identifying the electromagnetic counterparts to gravitational wave sources is vital to enabling the myriad of investigations possible with multimessenger astronomy. However, locating faint, fast-varying transients within large localisations remains challenging given the uncertainty in their detailed properties. In this work, we investigate how the nearby merger-induced GRBs would be localised by the LIGO–Virgo–KAGRA detector network during the fifth gravitational wave observing run (O5) and assess whether their optical counterparts could be detected using gravitational wave localisations alone, without additional localisation from gamma-ray instruments. Counterpart detectability is evaluated using the observed optical afterglow lightcurves of these GRBs and the distance-scaled lightcurve of the kilonova AT2017gfo as a fiducial template. We find that such events can be localised to comparatively small regions of the sky, often only a few to tens of square degrees. As a result, counterparts are detectable by at least one of the available optical telescopes during O5. However, detectability depends strongly on observational depth, as the counterparts are fainter than $22$ mag within a day. Facilities capable of reaching depths of $\gtrsim23$ mag therefore play a key role in recovering these faint counterparts. These results indicate that for such events during O5, the primary challenge for multimessenger discovery will be in achieving sufficient observational depth and reliably identifying the true counterpart among unrelated transients rather than gravitational wave localisation itself.

\end{abstract}


%%%%%%%%%%%%%%%%%%%%%%%%%%%%%%%%%%%%%%%%%%%%%%%%%%

%%%%%%%%%%%%%%%%% BODY OF PAPER %%%%%%%%%%%%%%%%%%
\section{Introduction}

Electromagnetic (EM) and gravitational wave (GW) observations of compact object mergers offer complementary data, shedding light on various aspects of these cosmic events. For instance, we can infer the masses and spins of the progenitors from GW detections, whereas EM detections can tell us about the mass and composition of ejecta, host galaxy associations, and underlying radiative processes \citep[e.g.,][]{metzger2020Kilonovae, margutti2021First}. The joint detection of both GW and EM signals plays a pivotal role in constraining the equation of state of neutron stars \citep[e.g.,][]{hinderer2010Tidal,
bauswein2013Prompt, abbott2018GW170817,
coughlin2018Constraints}, unravelling the origin of heavy elements generated in r-process nucleosynthesis \citep[e.g.,][]{just2015Comprehensive,
wu2016Production,
rosswog2017Detectability,
arcavi2017Optical,
pian2017Spectroscopic,
kilpatrick2017Electromagnetic,
kasliwal2022Spitzer}, and investigating the expansion rate of the universe by providing a new class of standard sirens \citep{schutz1986Determining, holz2005Using, dalal2006Short, nissanke2010EXPLORING, nissanke2013Determining, abbott2017Gravitationalwave}. All of this science relies on the identification of a counterpart that confirms the presence of ejecta, provides a precise position in the sky, and enables measurements of the source or host galaxy redshift. 

The mergers of compact objects offer a variety of routes for the production of luminous EM emission. Prior to the merger, it is possible that tidal forces on the neutron star give form to some kind of resonant shattering flare \citep{tsang2012Resonant}. During the merger, neutron-rich material is ejected in both tidal material and via a disk wind. This material can undergo rapid neutron captures, and the heating caused by the decay of the highly unstable massive isotopes powers a short-lived (days) faint ($M_V \sim -15$) transient now commonly referred to as a kilonova \citep{rosswog1998Mass,li1998Transient,metzger2010Electromagnetic}. Such kilonovae are expected to become very red due to heavy opacities from lanthanides created in the ejecta \citep[e.g.,][]{barnes2013Effect, tanaka2013Radiative, kasen2013OPACITIES}. Furthermore, in addition to the material launched at modest velocities in the kilonovae, it is likely that some of it is launched at ultra-relativistic velocities in the form of a jet, which creates both a gamma-ray burst normally expected to be of the short ($<2$s) variety, and an associated afterglow \citep{eichler1989Nucleosynthesis, narayan1992GammaRay, sari1998Spectra, lamb2019Optical, rossi2020Comparison}. 

Observations of short-duration gamma-ray bursts (GRBs) have generally supported this picture, with kilonovae identified in several events to date \citep[e.g.,][]{tanvir2013Kilonova, berger2013RProcess, gompertz2018Diversity, troja2019Afterglow}. The groundbreaking GW170817 \citep{abbott2017GW170817, abbott2017Multimessenger} provided the first unambiguous association between a short-GRB (GRB 170817A), its afterglow, and a kilonova (AT2017gfo) powered by r-process nucleosynthesis, firmly establishing binary neutron star (BNS) mergers as a progenitor channel for short-GRBs \citep{goldstein2017Ordinary, savchenko2017INTEGRAL, troja2017Xray, hallinan2017Radio, kasliwal2017Illuminating}. However, more recent observations of kilonovae accompanying long-duration ($>2$s) GRBs have challenged the traditional dichotomy between short and long-GRBs and pointed towards a broader range of progenitor channels for kilonovae, including neutron star–black hole (NSBH) mergers and magnetar-powered outflows \citep{rastinejad2022Kilonova, levan2024Heavyelement}. 

These observations imply that a number of GRBs with measured optical afterglows \citep{kann2010AFTERGLOWS, kann2011AFTERGLOWS}, and in some cases kilonovae, originate from compact object mergers. This naturally raises several questions: how would these events have been localised by the GW detectors had they been operating at the time? Could their optical counterparts, afterglow or kilonova, have been detected and identified using GW localisations alone? Addressing these questions can provide key insight into the prospects for joint GW–EM detections of such events in the upcoming fifth GW observing run O5\footnote{Updated observing capabilities available at \url{https://emfollow.docs.ligo.org/userguide/capabilities}}
 (2027+), including whether the primary bottleneck lies in GW detector sensitivity and localisation or in the depth and strategy of optical follow-up facilities. Additionally, it may also inform recommendations for follow-up campaigns.

 There have been several studies on predicting the detectability of EM counterparts for advanced ground-based GW detector networks and optical/infrared telescopes \citep[e.g.,][]{nissanke2011LOCALIZING, nissanke2013Identifying, colombo2023Multimessenger}. A commonly applied procedure in these studies is to synthesize a catalogue of BNS or NSBH events isotropically distributed within the reach of the target GW observing run, inject them into a network of detectors, select the events that result in a signal-to-noise ratio (SNR) above a certain threshold, and perform EM searches on the resulting localisations using GRB afterglow and kilonova brightness from different models. The models used to predict the counterpart brightness have several input parameters and rely on many simplifying assumptions. Hence, the predicted lightcurves might not be truly representative of the expected counterparts. These investigations are usually tailored to assess the detection capabilities of a specific survey under consideration \citep[e.g.,][]{arcavi2017Optical, frostig2022Infrared, mo2023Searching}. However, a few studies do consider a larger list of telescopes available at the time \citep[e.g.,][]{nissanke2013Identifying, chase2022Kilonova}. 

In this study, we compile a catalogue of past GRBs with a plausible merger origin at nearby redshifts ($z < 0.15$) and investigate if we would have detected their optical counterparts (without any aid from the gamma-ray instruments like \textit{Swift} or \textit{Fermi}) had GW detectors been operating at their projected O5 sensitivity when these events occurred. To do this, each event is placed at its observed luminosity distance and sky position, and a fixed set of fiducial source parameters is adopted to compute the corresponding GW signal and localisation. What distinguishes our study is the utilization of actual afterglow brightness data rather than model-based predictions, thereby minimizing uncertainties in the detectability limits we derive. To gain complementary insight into kilonova detectability, we use the distance-scaled lightcurve of AT2017gfo as a fiducial template and assess whether it would be detectable in isolation for the events in our sample. For simplicity, we adopt this fiducial lightcurve for all events and do not use the kilonova candidates reported for some of our GRBs. It is worth noting that our analysis utilizes on-axis events, which are likely to have smaller GW localised error regions and their afterglows exhibit higher peak brightness compared to their off-axis analogues, therefore, should be taken as an optimistic scenario.

This paper is organised as follows. In Section~\ref{sec: methods} we describe the GRB sample, the counterpart lightcurves, and the simulations used to generate GW skymaps and optical follow-up observations. The resulting skymaps and counterpart detectability are presented in Section~\ref{sec: results}. In Section~\ref{sec: discussions}, we discuss the implications of these results for multimessenger detections and follow-up strategies during O5. Our conclusions are summarised in Section~\ref{sec: conclusions}.

\section{Methods}\label{sec: methods}

\subsection{GRB Sample and Counterpart Lightcurve}

\subsubsection{GRB Sample Selection}\label{subsec: grb-data}
\begin{table}
\setlength\extrarowheight{1pt}
\centering
\caption{GRBs used in this study along with their durations ($T_{90}$), luminosity distances ($D_L$), celestial coordinates - right ascension (RA) and declination (Dec), and references for their optical afterglow data.}
\label{tab:grb-sample}
\begin{threeparttable}
\begin{tabular}{cccccc}
\toprule
GRB Name & \thead{$T_{90}$\\(s)} & \thead{$D_L$\\(Mpc)} & \thead{RA\\(deg)} & \thead{Dec\\(deg)} & \thead{Afterglow \\References} \\ \midrule
060505 & 4 & 411 & 331.76 & -27.81 & \textcolor{black}{a}\\ 
061201 & 0.8 & 518 & 332.13 & -74.58 & b\\
060614 & 109  & 589 & 320.88 & -53.02 & c\\
080905A & 1 & 573 & 287.67 & -18.88 & d\\
150101B & 0.01 & 635 & 188.02 & -10.93 & e\\ 
211211A & 136.5 & 346 & 212.29 & 27.88 & f\\ 
230307A & 35 & 294 & 60.85 & -75.37 & g\\ 
\bottomrule
\end{tabular}
\begin{tablenotes}
% \item[\dag] Luminosity Distance\\ \tnote{\dag}
a. \citet{xu2009SEARCH, ofek2007GRB, kann2011AFTERGLOWS}
b. \citet{stratta2007Study, kann2011AFTERGLOWS} 
c. \citet{yang2015Possible, dellavalle2006Enigmatic, gal-yam2006Novel, xu2009SEARCH, kann2011AFTERGLOWS}
d. \citet{rowlinson2010Discovery, pagani2008Swift, nakajima2008GRB, kann2011AFTERGLOWS}
e. \citet{fong2016AFTERGLOW} 
f. \citet{rastinejad2022Kilonova}
g. \citet{levan2024Heavyelement} 
\end{tablenotes}
\end{threeparttable}
\end{table}

For our analysis, we select \textit{Swift} and \textit{Fermi} GRBs that meet two criteria. First, each burst has a reasonably secure redshift of $z < 0.15$. This redshift is motivated by the sky-averaged detection range expected\footnote{Calculated using the python notebook made available by \citet{chen2021Distance}, adopting the same O5 noise curves used in our GW simulations (see footnote~\ref{fn:psd}).} for the EM-bright NSBH mergers during O5, with the LIGO–Virgo–KAGRA (LVK) Network operating at the higher end of design sensitivity. Second, the burst is either a short-GRB ($T_{90} < 2$ s) or a long-GRB ($T_{90} > 2$ s) with a plausible merger origin, i.e., either accompanied by a kilonova or lacking an associated supernova. We end up with three long-GRBs and four short-GRBs ranging between 294 and 635 Mpc as summarized in Table \ref{tab:grb-sample}.

\subsubsection{GRB Afterglow Data}
\begin{figure}
\includegraphics[width=\columnwidth]{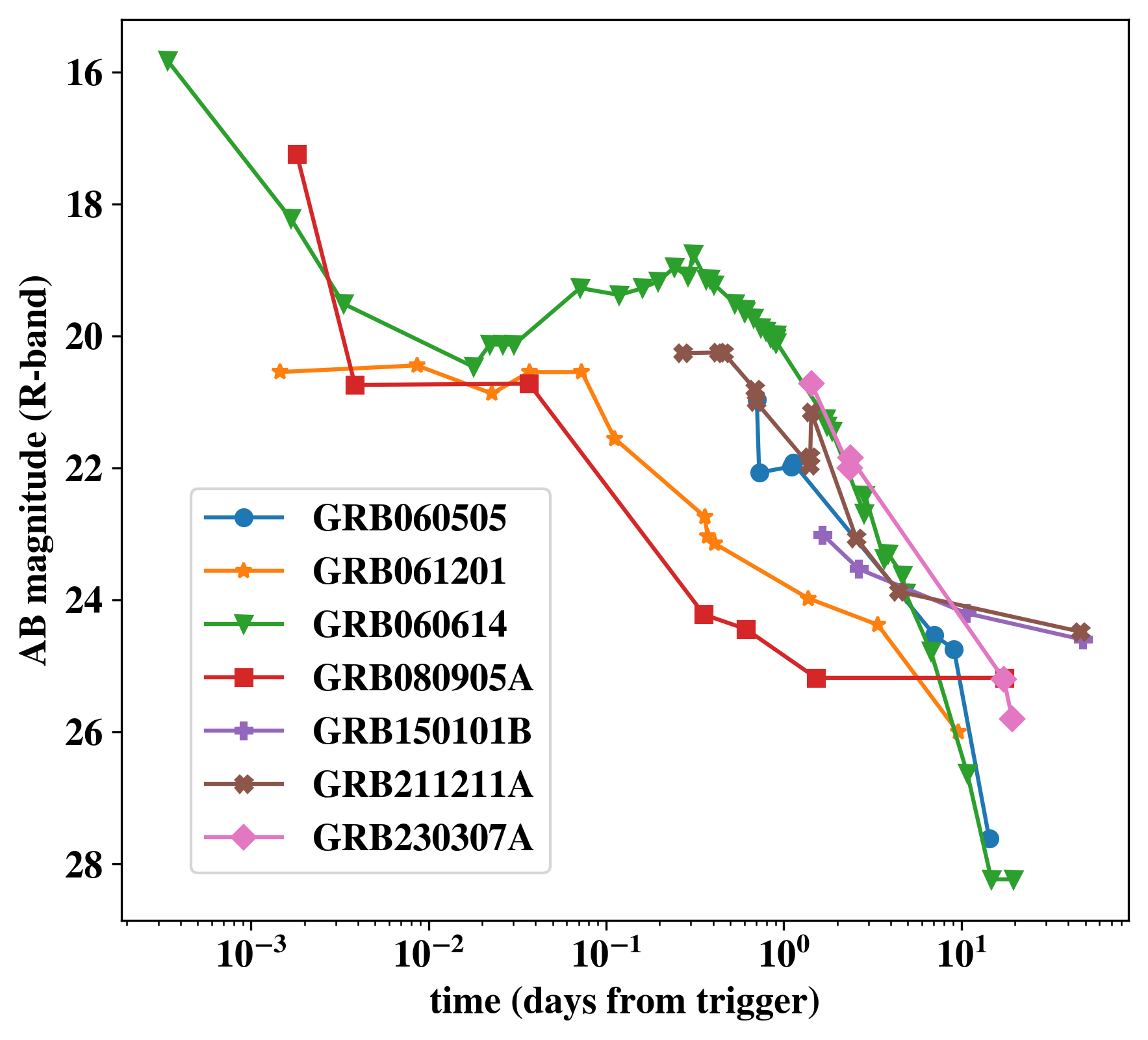}
    \caption{Optical afterglows of GRBs in our dataset. The AB mag in the R-band is plotted on the vertical axis, while time since GRB trigger is on the horizontal axis. Each GRB is represented by a distinct color and marker.}
    \label{fig:afterglow-data}
\end{figure}

Figure \ref{fig:afterglow-data} shows the optical afterglow lightcurves for these GRBs, compiled from observations reported in the references listed in Table \ref{tab:grb-sample}. We treat measurements in the $r$, $R$, and $R_c$ bands as equivalent, without applying detailed colour-term corrections. These lightcurves are then compared to the $R$-band limiting magnitudes of the telescopes when assessing detectability. We do this both because such terms depend on the underlying spectral energy distribution of the source, which is not always well known, and because small offsets in the measurements do not substantially impact our outcomes. 

\subsubsection{Kilonova Lightcurve Template}
\label{subsec: kilonova-template}

In addition to the observed afterglow data, we incorporate the $R$-band lightcurve of AT2017gfo, scaled to the luminosity distance of each GRB, to assess kilonova detectability independently. Although several GRBs in our sample have reported kilonova candidates, these events are typically sparsely sampled and require subtraction of a bright afterglow component, making their intrinsic kilonova lightcurves uncertain \citep[see e.g.,][]{tanvir2013Kilonova, berger2013RProcess, gompertz2018Diversity}. AT2017gfo, by contrast, is the best-sampled kilonova to date, with densely sampled multi-band photometry spanning the optical and near-infrared \citep[see e.g.,][]{smartt2017Kilonova, kasliwal2017Illuminating, villar2017Combined}. Its optical luminosity, while somewhat brighter, lies broadly within the range inferred for other claimed kilonovae associated with GRBs. We therefore adopt AT2017gfo as a fiducial template without attempting to model event-to-event diversity in ejecta mass, opacity, or viewing angle. The lightcurve is used only within the temporal range covered by the AT2017gfo data and without assuming a specific functional form for its temporal evolution. Because all events in our sample lie at $z<0.15$, we do not apply explicit K-corrections and treat the observed lightcurve as approximately representative of the observed-frame $R$-band emission at these redshifts.

\subsection{Simulating GW Skymaps}

To determine astrophysically realistic skymaps resulting from GW alerts for each GRB, we simulate GW strain for these events using the \texttt{IMRPhenomPv2} \citep{hannam2014Simple, husa2016Frequencydomain, khan2016Frequencydomain} waveform model, inject them into Gaussian noise coloured by the power spectral density (PSD) of different GW detectors, and perform a full parameter estimation on the resulting data stream for a network of detectors. We do this for two different progenitor assumptions -- BNS and NSBH. At zeroth order, tidal deformabilities of the neutron stars do not appear to significantly affect the GW-inferred sky localisations \citep[e.g.,][]{singer2016Rapid}. Hence, for the purpose of this study, we choose to neglect neutron star tidal effects and treat both classes of compact objects as point masses, adopting a waveform model without tidal corrections such as \texttt{IMRPhenomPv2}. Ignoring the tidal deformabilities, the GW strain of a quasi-circular BNS/NSBH system can be characterised by 15 parameters -- component masses ($m_1, m_2$), spin magnitudes ($\chi_1, \chi_2$), four angles specifying the spin orientations ($\theta_1, \theta_2, \phi_{12}, \phi_{jl}$), polarization angle ($\psi$), time of coalescence ($t_c$), orbital phase at coalescence ($\phi_c$), luminosity distance ($D_L$), inclination angle ($\iota$, defined as the angle between the orbital angular momentum and the line of sight), right ascension (RA), and declination (Dec) \citep{cutler1994Gravitational}. We use the LVK network (H1, L1, V1, and K1) with PSDs\footnote{\label{fn:psd}Corresponding files \texttt{AplusDesign.txt}, \texttt{avirgo\textunderscore O5high\textunderscore NEW.txt}, and \texttt{kagra\textunderscore 128Mpc.txt} can be found in \url{https://dcc.ligo.org/LIGO-T2000012/public}} corresponding to the higher end of planned sensitivity during O5 \citep{abbott2020Prospects}, which makes our simulations an optimistic O5 scenario.

\begin{table}
\centering
\setlength\extrarowheight{1pt}
\begin{tabular}{ccc}
\hline
\textbf{Parameter} & \textbf{Fixed?} & \textbf{Notes} \\
\hline
Masses ($m_1, m_2$) & Yes & NS: 1.4 $M_\odot$, BH: 5 $M_\odot$\\
Spin magnitudes ($\chi_1, \chi_2$) & Yes & Zero \\
Spin orientations & Yes & Zero (non-precessing) \\
Inclination angle ($\iota$) & Yes & Zero (face on) \\
Luminosity distance ($D_L$) & No & From GRB data\\
Right Ascension (RA) & No & From GRB data\\
Declination (Dec) & No & From GRB data\\
Coalescence time ($t_c$) & No & From GRB data\\
Polarization angle ($\psi$) & No & Random per event \\
Coalescence phase ($\phi_c$) & No & Random per event \\
\hline
\end{tabular}
\caption{Parameters used to generate the injected GW signals for the simulated events. The table indicates whether each parameter is fixed across the sample or taken from the observed properties of the individual GRBs.}
\label{tab:parameters}
\end{table}

For simulating the GW strain signal, we adopt the source parameters summarised in Table~\ref{tab:parameters}, which lists the injected parameters and indicates whether they are fixed or taken from the observed GRB properties. In particular, we assume zero spin for both components, corresponding to non-precessing systems. We adopt component masses of 1.4$M_\odot$--1.4$M_\odot$ for a BNS and 1.4$M_\odot$--5$M_\odot$ for an NSBH. In the NSBH case, the choice of zero BH spin motivates adopting a 5$M_\odot$ BH mass, ensuring that the BH’s innermost circular orbit (ICO) is sufficiently small relative to the tidal disruption radius to allow disruption of the NS companion before plunge. However, note that a different choice of BH spin, particularly a prograde spin aligned with the orbital angular momentum, can make higher BH masses viable choices for tidal disruption \citep[see e.g.,][]{nissanke2013Identifying, foucart2018Remnant}. The GRBs detected by \textit{Swift}/\textit{Fermi} are typically viewed face on along the jet, so the inclination angle of their binary progenitor ($\iota$) would be a small angle, but for simplicity, we use $\iota$ = 0. For each event, we fix the coalescence phase and polarisation angle to arbitrary values within the intervals $[0,2\pi]$ and $[0,\pi]$, respectively, and use the same values consistently across both progenitor cases. The GRB trigger times are used as the coalescence times. The luminosity distance, right ascension, and declination are fixed to the observed values of the individual GRBs. It is worth noting that parameters derived directly from the GRB observations consist of five quantities: the 3D sky position (D$_L$, RA, and Dec), the event time, and the inclination angle. The remaining ten parameters are fiducially set based on various assumptions described above. 

We perform a full parameter estimation on the simulated GW data with the \texttt{Bilby} software \citep{ashton2019Bilby} using the reduced order quadrature (ROQ) technique \citep{smith2016Fast}. Our analysis uses data segments in the frequency range 32-2048 Hz sampled at 4096 Hz. A 32 Hz low-frequency cutoff ensures that the inspiral of low-mass systems, such as a 1.4$M_\odot$--1.4$M_\odot$ BNS, lies within the validity range of the ROQ basis used\footnote{See the IMRPhenomPv2 128s ROQ basis data available at \url{https://git.ligo.org/lscsoft/ROQ_data}} \citep{smith2016Fast}. Although the GW strain data was generated using the observed GRB properties, each event is analyzed as a standard GW detection without external localisation or distance information, so that the resulting skymaps reflect what would be obtained from the GW data alone. Hence, we adopt the standard astrophysically motivated priors used in GW inference. Specifically, we assign uniform priors to the chirp mass and geocentric coalescence time, with widths of 0.2 $M_\odot$ and 0.2 s centered on the injected values. The mass ratio is sampled uniformly over [0.125, 1]. We adopt uniform priors on spin magnitudes over [0, 0.05] for neutron stars and [0, 0.80] for black holes. The low-spin prior for neutron stars is motivated by observations of Galactic BNS systems \citep[e.g.,][]{lorimer2008Binary} and GW170817 \citep{abbott2019Properties}, while the upper bound on BH spin is set by the validity range of the ROQ basis. For the luminosity distance sampling, we adopt a power-law prior over the range 1–1000 Mpc, corresponding to a source population distributed uniformly in volume. The remaining parameters are assigned standard priors prescribed in \citet{ashton2019Bilby, romero-shaw2020Bayesian}.

\subsection{Simulating Observation Schedules}\label{subsec: sim_observation_schedule}

For the purposes of our EM follow-up simulations, we do not impose the GW detection threshold (network SNR > 12). Events with network SNR below the nominal detection limit are still included, even if they would not be formally reported as detections in GW alerts. Our goal is to explore how different skymaps could be searched in practice, rather than restricting the analysis to only formally detected GW events.

We compile a list of telescopes that are operational and have resources that can be allocated for GW follow-up in O5, namely Zwicky Transient Facility (ZTF; \citealt{bellm2018Zwicky}), Dark Energy Camera (DECam; \citealt{flaugher2015DARK}), Asteroid Terrestrial-impact Last Alert System (ATLAS; \citealt{tonry2018ATLAS}), Gravitational-wave Optical Transient Observer (GOTO; \citealt{gompertz2020Searching}), Panoramic Survey Telescope and Rapid Response System (Pan-STARRS; \citealt{morgan2012Design}), BlackGEM \citep{bloemen2015BlackGEM, groot2024BlackGEM}, Vera C. Rubin Observatory (Rubin; \citealt{ivezic2019LSST}) and Nancy Grace Roman Space Telescope (Roman; \citealt{spergel2015WideField}). Among these, we treat telescope units at different geographical locations as different telescopes, even if they are part of the same survey. However, in practice, these surveys might schedule observations in a coordinated fashion. Meanwhile, some of the telescopes, namely GOTO, Pan-STARRS, and BlackGEM, also have multiple (say $N$) units located in distinct domes at the same geographical site. We assume these individual units operate in a coordinated fashion to cover the searchable area $N$ times faster, rather than choosing to observe the same field with different filters simultaneously. Table \ref{tab:telescopes} enumerates all the telescopes, their exposure times, readout overheads, and limiting magnitudes used in our analysis. For each telescope, we consider shallow and deep exposures. The indicated exposure times in both cases represent nominal values employed by these telescopes, with actual exposure times subject to variation based on the specific observing strategy, prevailing seeing conditions, and other factors at the time of observation. The limiting magnitudes listed are the 5$\sigma$ depths achieved by these telescopes with the specified exposures during dark times. 

Subsequently, we create an observation schedule for skymaps of all GRBs in our sample for two different BNS and NSBH progenitor scenarios. We employ the ranked tiling strategy \citep{ghosh2016Tiling} to schedule observations\footnote{We use a modified version of the original code, which can be found in \url{https://github.com/krishnakruthi/sky_tiling}}. Here, for each telescope, the whole sky is divided into a fixed grid of tile size determined by the telescope's field-of-view (FOV). The tiles are ranked based on their localisation probabilities calculated from the skymaps, and tiles with the highest ranks are selected such that they cover a 90\% credible area. Then, observations are scheduled based on their visibility (altitude > 30$^\circ$) and rank, covering a maximum of 50 hours of observations spanning multiple nights. Note that our simulation does not account for telescope slew rates. We assume an optimistic latency of 15 minutes between the event trigger time and the start of observations, except in the case of Roman, where it is set to 2 hours to allow for the delays in Target of Opportunity (ToO) triggers. Surveys typically require at least two images of the same field to confirm the presence of a transient source and to reject unrelated sources and detector artifacts. In practice, these images may be separated by tens of minutes. For simplicity, we assume that each telescope pointing consists of two consecutive exposures of the same field taken back-to-back.

\begin{table*}
\centering
\begin{threeparttable}
\caption{An overview of telescopes utilized in our analysis, including their location, latitude, field of view (FOV), exposure times (shallow and deep), readout overhead, and the 5$\sigma$ limiting AB magnitude in R-band for respective exposure times. References for each telescope's specifications are also provided.}
\label{tab:telescopes}
\begin{tabular}{cccccccccc}
\toprule
Telescope &
  Location &
  Latitude &
  \thead{Units\\/Domes} &
  \thead{FOV \\ ($\mathrm{deg^2}$)} &
  \thead{Exposure (s) \\ (Shallow | Deep)} &
  \thead{Overhead (s)} &
  \thead{5$\sigma$ depth\tnote{c} \\ (Shallow | Deep) } & Ref. \\ \midrule
ZTF             & Palomar, California  & 33.36\degree \ N & 1 & 47  & 30 | 300 & 10  & 20.9 | 22.2 & i\\
DECam (DES)     & Cerro Tololo, Chile & 30.17\degree \ S & 1 & 3   & 40 | 600 & 20   & 23.3 | 24.8 & ii\\
ATLAS-Sutherland & SAAO, Sutherland\tnote{b} & 32.38\degree \ S & 1 & 30 & 30 | 600 & 6 & 19.3 | 20.9 & iii\\
ATLAS-Maunaloa  & Mauna Loa, Hawaii & 19.54\degree \ N & 1 & 30  & 30 | 600 & 6    & 19.3 | 20.9 & iii\\
ATLAS-Haleakala & Haleakala, Hawaii & 20.71\degree \ N & 1 & 30  & 30 | 600 & 6 & 19.3 | 20.9 & iii\\
ATLAS-Chile     & El Sauce, Chile  & 30.47\degree \ S & 1 & 30  & 30 | 600 & 6    & 19.3 | 20.9 & iii\\
GOTO-16         & La Palma, Spain & 28.71\degree \ N & 2 & 40  & 60 | 300 & 10   & 18.7 | 19.6 & iv\\
GOTO-16         & Siding Spring, Australia & 31.27\degree \ S & 2 & 40  & 60 | 300 & 10   & 18.7 | 19.6 & iv\\
Pan-STARRS      & Haleakala, Hawaii & 20.71\degree \ N & 2 & 7   & 45 | 600 & 10.3 & 21.8 | 23.2 & v\\
BlackGEM        & La Silla, Chile & 29.26\degree \ S & 3 & 2.7 & 60 | 300 & 25   & 21.1 | 22.0 & vi\\
Rubin & Cerro Pachón, Chile & 30.24\degree \ S & 1 & 9.6 & 30 | 300 & 2 & 24.4 | 25.6 & vii\\
Roman & Sun–Earth L$_2$ & Space-based & 1 & 0.28 & 55 | 300 & 71 & 25.6 | 27.2 & viii\\
\bottomrule
\end{tabular}
\begin{tablenotes}[item]
\item[b] SAAO: South African Astronomical Observatory 
\item[c] limiting AB magnitude in R-band
\end{tablenotes}
\vspace{1mm}
\begin{itemize}
\item[(i)] ZTF: The exposure time and depth are motivated by \citet{ahumada2022Search}; we take the median readout overhead from \url{https://www.ztf.caltech.edu/ztf-camera.html}.
\item[(ii)] DECam: The exposure time is motivated by \citet{garcia2020DESGW}, and the corresponding 5$\sigma$ limiting magnitude is calculated using the formula in \citet{bom2023Designing}. The readout time is taken from \url{https://noirlab.edu/science/programs/ctio/instruments/Dark-Energy-Camera/characteristics}
\item[(iii)] ATLAS: The exposure time is motivated by \citet{smartt2023GW190425}. The corresponding R-band depth is taken from \url{https://atlas.fallingstar.com/specifications.php}
\item[(iv)] GOTO: \citet{dyer2020Telescope}; \textcolor{black}{readout is arbitrarily set to 10s}
\item[(v)] Pan-STARRS: The exposure times are motivated by \citet{smartt2023GW190425}. The corresponding 5$\sigma$ limiting magnitude and readout time are taken from \citet{chambers2019PanSTARRS1}.
\item[(vi)] BlackGEM: \citet{groot2024BlackGEM}
\item[(vii)] Rubin: We use fiducial exposure time from \citet{cowperthwaite2019LSST}. The corresponding limiting magnitude is calculated using information available on \url{https://smtn-002.lsst.io/}
\item[(viii)] Roman: \url{https://roman.gsfc.nasa.gov/science/WFI_technical.html}
\end{itemize}
\end{threeparttable}
\end{table*}

\subsection{Counterpart Detectability Criteria}\label{sec: detection-criteria}

For each event, we search the observation schedule for the source tile. If found, we classify an observation of the source tile with a given telescope as a detection if the estimated counterpart brightness exceeds the nominal 5$\sigma$ limiting magnitude of the telescope, allowing for an effective uncertainty of $\pm$0.5 mag. This tolerance is introduced to account for variations in observing conditions, such as lunar illumination or airmass, and uncertainties in the counterpart lightcurves described below. Moreover, our simulation does not attempt to model the detailed transient-detection criteria employed by individual surveys. Instead, we assume that the availability of two consecutive images per pointing (as described in Section~\ref{subsec: sim_observation_schedule}) is sufficient for a survey to detect the source.

For the afterglows, the source brightness at the time of observation is estimated by linearly interpolating between adjacent afterglow data points or extrapolating the lightcurve using a power-law decay for epochs outside the available dataset, as GRB afterglows are known to follow power-law decays. The adopted $\pm$0.5 mag tolerance accounts for uncertainties associated with interpolation and extrapolation of the observed lightcurves, approximate band matching between $r$, $R$, and $R_c$ measurements (see Section~\ref{subsec: grb-data}), and the measurement uncertainties of the data points. We adopt the same observed GRB afterglow data for both BNS and NSBH progenitor scenarios, assuming identical afterglow brightness.

For the kilonovae, as described in Section \ref{subsec: kilonova-template}, we do not apply explicit K-corrections, and the same $\pm$0.5 mag tolerance is applied when comparing the scaled AT2017gfo brightness to telescope limiting magnitudes. However, this margin does not attempt to capture the full intrinsic diversity of kilonova luminosities.

Furthermore, we emphasize that detection alone does not constitute a discovery. To claim a discovery and securely identify the counterpart, multiple detections of the source tile separated in time are required in order to distinguish the afterglow from unrelated transients and variable sources. In practice, potential candidates for the counterpart are selected based on information about their host galaxy, as well as their colour and magnitude evolution. Only after a thorough photometric and spectroscopic follow-up of these candidates, the counterpart is successfully identified \citep[e.g.,][]{nissanke2013Identifying, abbott2017Multimessenger, rastinejad2021Probing, ahumada2022Search}.

\section{Results} \label{sec: results}
\subsection{Skymaps}\label{sec: results-skymaps}
\begin{figure*}
\includegraphics[width=\textwidth]{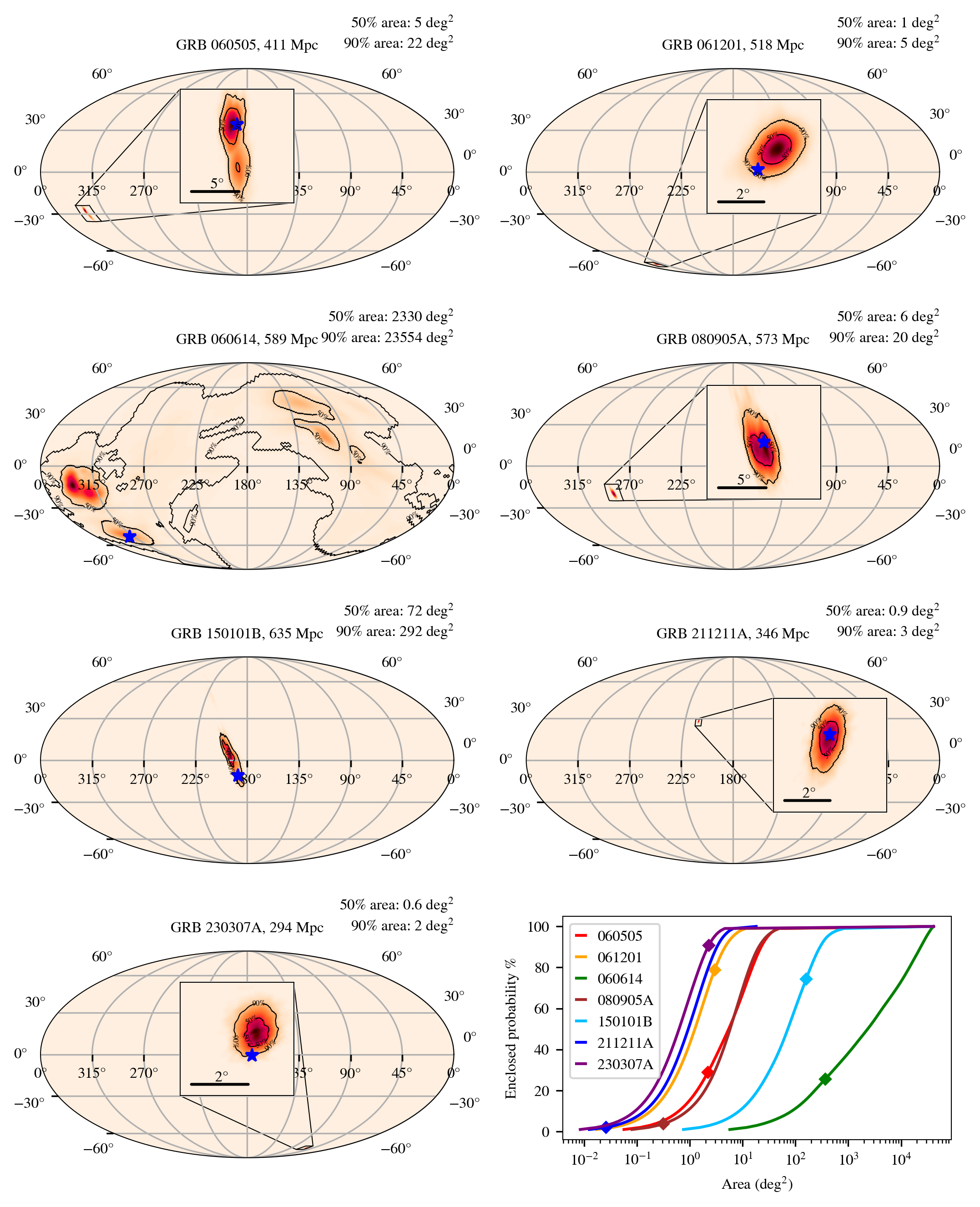}
\caption{BNS O5 skymaps: simulated skymaps showing the 50\% and 90\% credible regions obtained using \texttt{Bilby} for the BNS progenitor scenario with GW detectors (LIGO Hanford, LIGO Livingston, Virgo, and KAGRA) operating at O5 sensitivity. The location of the injected source is marked by a blue star. Bottom right: Plot showing enclosed probability percentage for a given area for each of the events.}
\label{fig:BNS-O5-skymaps}
\end{figure*}

\begin{figure*}
\includegraphics[width=\textwidth]{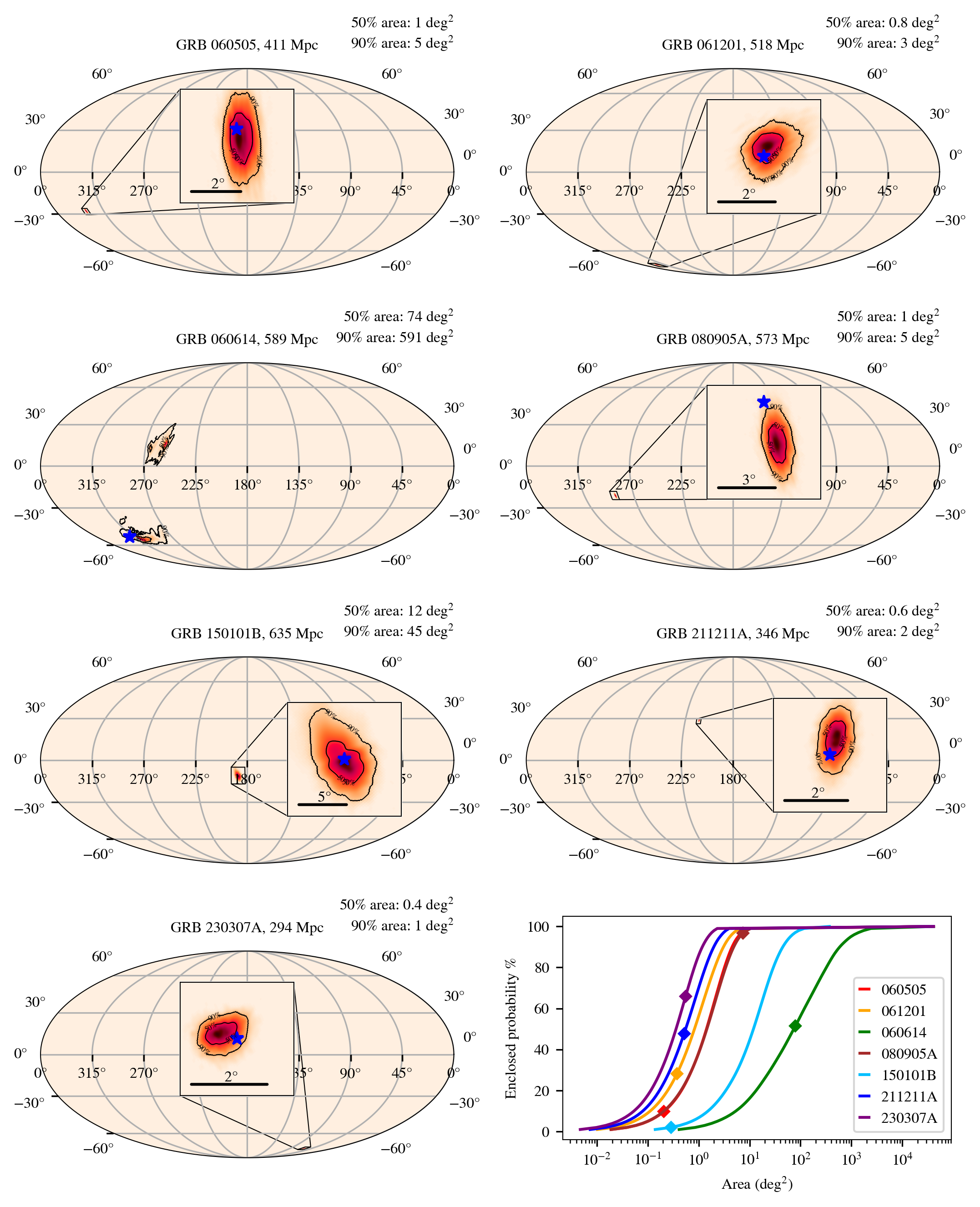}
\caption{NSBH O5 skymaps: simulated skymaps showing the 50\% and 90\% credible regions obtained using \texttt{Bilby} for the NSBH progenitor scenario with GW detectors (LIGO Hanford, LIGO Livingston, Virgo, and KAGRA) operating at O5 sensitivity. The location of the injected source is marked by a blue star. Bottom right: Plot showing enclosed probability percentage for a given area for each of the events.}
\label{fig:NSBH-O5-skymaps}
\end{figure*}

We present the GW skymaps resulting from our simulations in Figures \ref{fig:BNS-O5-skymaps} and \ref{fig:NSBH-O5-skymaps}, and summarise the corresponding network SNRs and sky localisations in Table \ref{tab:gw-statistics}. For the 7 events in our simulations, the BNS progenitor case yields three events (43\%) that are very well localised, with areas $\leq$ 5 deg$^2$; two (29\%) with an intermediate localisation of 20 and 22 deg$^2$, comparable to GW170817; one event (14\%) with a moderately large localisation of 292 deg$^2$; and one (14\%) poorly localised event, with an area exceeding 20000 deg$^2$. Four events (57\%) have network SNRs above the nominal detection threshold (network SNR $\geq$ 12), while three (43\%) fall in what we define here as the sub-threshold regime (7 $\leq$ network SNR $\leq$ 12). On the other hand, in the NSBH progenitor case, five events (71\%) are very well localised with areas $\leq$ 5 deg$^2$; one event (14\%) shows intermediate localisation of 45 deg$^2$; and one event (14\%) has a moderately large localisation of 591 deg$^2$. Here, six events (86\%) have network SNRs above the detection threshold while only one event (14\%) is subthreshold. No simulated events in either case have network SNR $<$ 7. While these numbers represent sky localisations that could be realistically expected for the events in our sample, it should be emphasized that they correspond to an optimistic configuration of a four-detector LVK network operating at the higher end of design sensitivity for O5.

Notably, our analysis reveals a few deviations from simple SNR- or distance-based expectations for sky localisation (see Table \ref{tab:gw-statistics}). Firstly, events at comparable distances can exhibit markedly different SNRs and sky localisations. GRB 060614 (589 Mpc) and GRB 080905A (573 Mpc), for instance, have BNS network SNRs of 7.3 and 11.1, corresponding to localisation areas of 23554 deg$^2$ and 20 deg$^2$, respectively. Secondly, even for nearly identical SNRs, localisation areas can differ substantially. For example, GRB 060505 and GRB 061201 have network SNRs of 13.3 and 13.1 in the BNS scenario, yet their sky areas are 22 deg$^2$ and 5 deg$^2$ respectively (a factor of 4.4 improvement despite the slightly lower network SNR). Lastly, the assumption that more distant events are more poorly localised does not always hold. For instance, GRB 060505 at 411 Mpc has a localisation of 22 deg$^2$ in the BNS scenario, while the more distant GRB 061201 at 518 Mpc is localised to only 5 deg$^2$. These examples highlight the strong influence of network geometry and source sky position on GW localisations.

Under the NSBH assumption, the sky localisations for two previously poorly localised events change substantially compared to the BNS assumption. For GRB 060614, the localisation area changes from 23554~deg$^2$ (BNS) to 591~deg$^2$ (NSBH), corresponding to a factor of $\sim$40 improvement, while for GRB 150101B it changes from 292~deg$^2$ (BNS) to 45~deg$^2$ (NSBH), a factor of $\sim$6.5 improvement. For the two events with localisation areas comparable to GW170817, the sky areas change from 20-22 deg$^2$ to 5 deg$^2$ in both cases, corresponding to a factor of $\sim$4 improvement.

The enclosed probability distributions as a function of sky localisation area (bottom-right panels of Figures~\ref{fig:BNS-O5-skymaps} and \ref{fig:NSBH-O5-skymaps}) exhibit a gradual rise at low probabilities, a steeper increase at intermediate values, and flatten as the enclosed probability approaches unity. Consequently, the 90\% credible regions are typically larger by a factor of $\sim$3--4 than the corresponding 50\% credible regions. For instance, in the case of GRB~150101B, the 90\% and 50\% areas are 291.8~deg$^2$ and 71.6~deg$^2$, respectively, for a BNS progenitor, and 44.9~deg$^2$ and 11.9~deg$^2$ for an NSBH progenitor. This difference is even more pronounced for the worst-localised event, GRB~060614, where the 90\% and 50\% areas are 23554~deg$^2$ and 2330~deg$^2$, respectively, in the BNS scenario, and 591~deg$^2$ and 74~deg$^2$ for the NSBH case. See Table~\ref{tab:gw-statistics} for the corresponding network SNRs of these events. Furthermore, on average, the true source locations lie within the 44\% credible regions of their respective skymaps. Across both BNS and NSBH scenarios, a total of 8 out of 14 events ($\sim$57\%) have their true source locations within the 50\% credible regions, 10 out of 14 ($\sim$71\%) within the 75\% regions, and 12 out of 14 ($\sim$86\%) within the 90\% regions, consistent with statistical expectations given the limited sample size. Here, the two outliers are GRB~230307A in the BNS scenario, with the true source at the 91\% credible level, and GRB~080905A in the NSBH scenario, at 97\% credible level.

\begin{table*}
\centering
\setlength\extrarowheight{2pt}
\begin{threeparttable}
\caption{Network SNRs and sky localisations for GW simulations of GRB events in our sample during the O5 observing run for two different progenitor scenarios - BNS and NSBH. The O5 network consists of H1, L1, V1 and K1.}
\label{tab:gw-statistics}
\begin{tabular*}{\textwidth}{@{\extracolsep{\fill}}*{7}{c}}
\toprule
\multirow{2}{*}{GRB Name} & \multirow{2}{*}{$D_L$ (Mpc)} & \multicolumn{2}{c}{BNS} & \multicolumn{2}{c}{NSBH} \\ 
\cline{3-4} \cline{5-6}
& & Network SNR & Area\tnote{a} ($\mathrm{deg^2}$) & Network SNR & Area\tnote{a} ($\mathrm{deg^2}$) \\ \midrule
060505   & 411 & 13.3 & 22     & 20.9 & 5   \\
060614   & 589 & 7.3  & 23554  & 11.5 & 591 \\
061201   & 518 & 13.1 & 5      & 20.7 & 3   \\
080905A  & 573 & 11.1 & 20     & 17.5 & 5   \\
150101B  & 635 & 8.7  & 292    & 13.8 & 45  \\
211211A  & 346 & 20.8 & 3      & 32.8 & 2   \\
230307A  & 294 & 23.0 & 2      & 36.3 & 1   \\
\bottomrule
\end{tabular*}

\begin{tablenotes}[item]
\item[a] 90\% credible region 
\end{tablenotes}
\end{threeparttable}
\end{table*}

\subsection{Detectability}\label{sec: results-detectability}

\begin{figure*}
\includegraphics[width=0.85\textwidth]{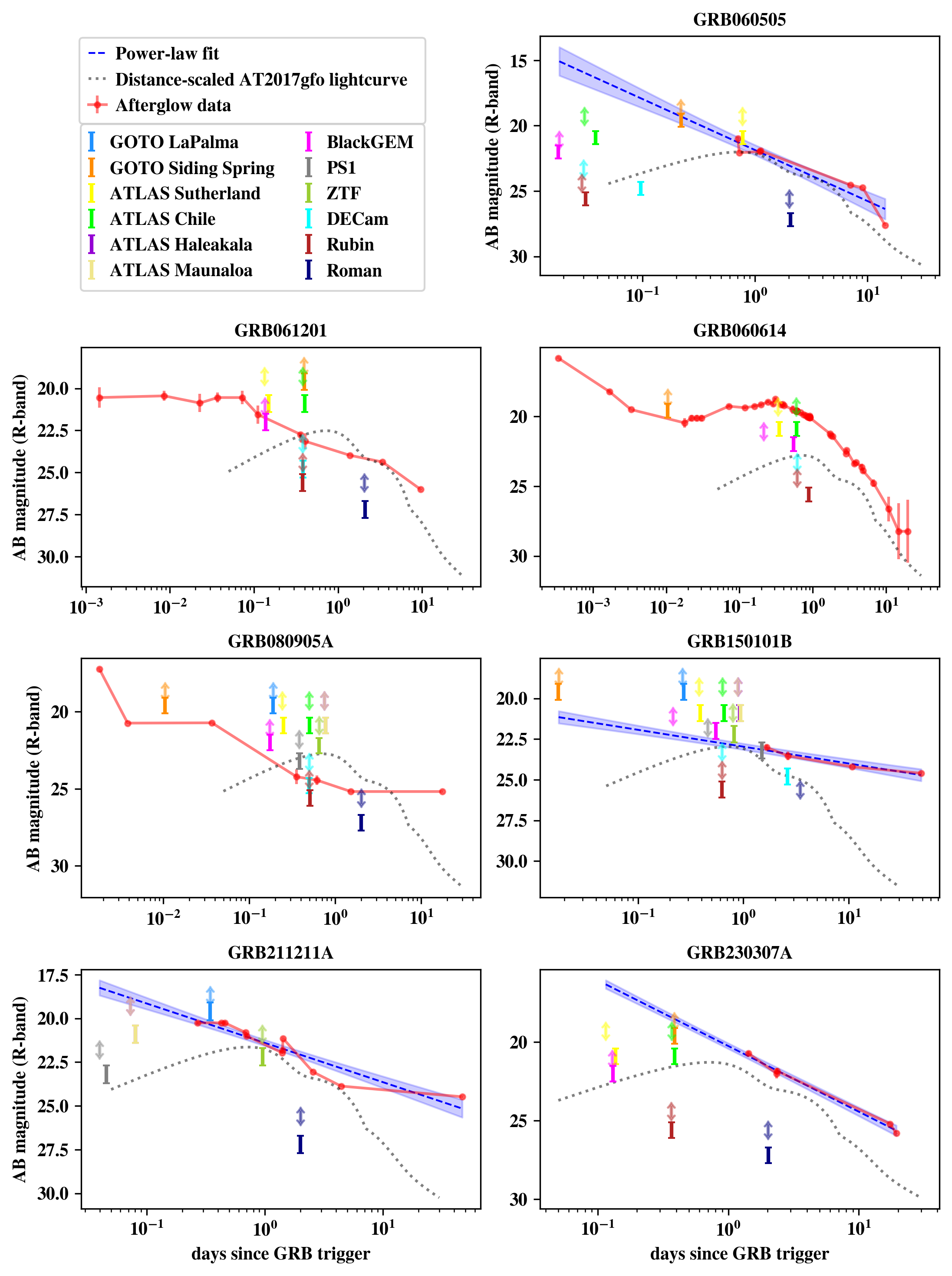}
    \caption{BNS Detectability: The afterglow lightcurves of GRBs with double-arrow (rook-shaped) markers representing the time of source-tile observation and the limiting magnitude of telescopes for shallow (deep) searches in an O5 scenario with BNS progenitor. The markers are color-coded to represent different telescopes, as shown in the plot legend. The shaded blue regions show the uncertainty associated with power-law extrapolation of the afterglow lightcurves.}
    \label{fig:BNS_o5_detectability}
\end{figure*}

\begin{figure*}
\includegraphics[width=0.85\textwidth]{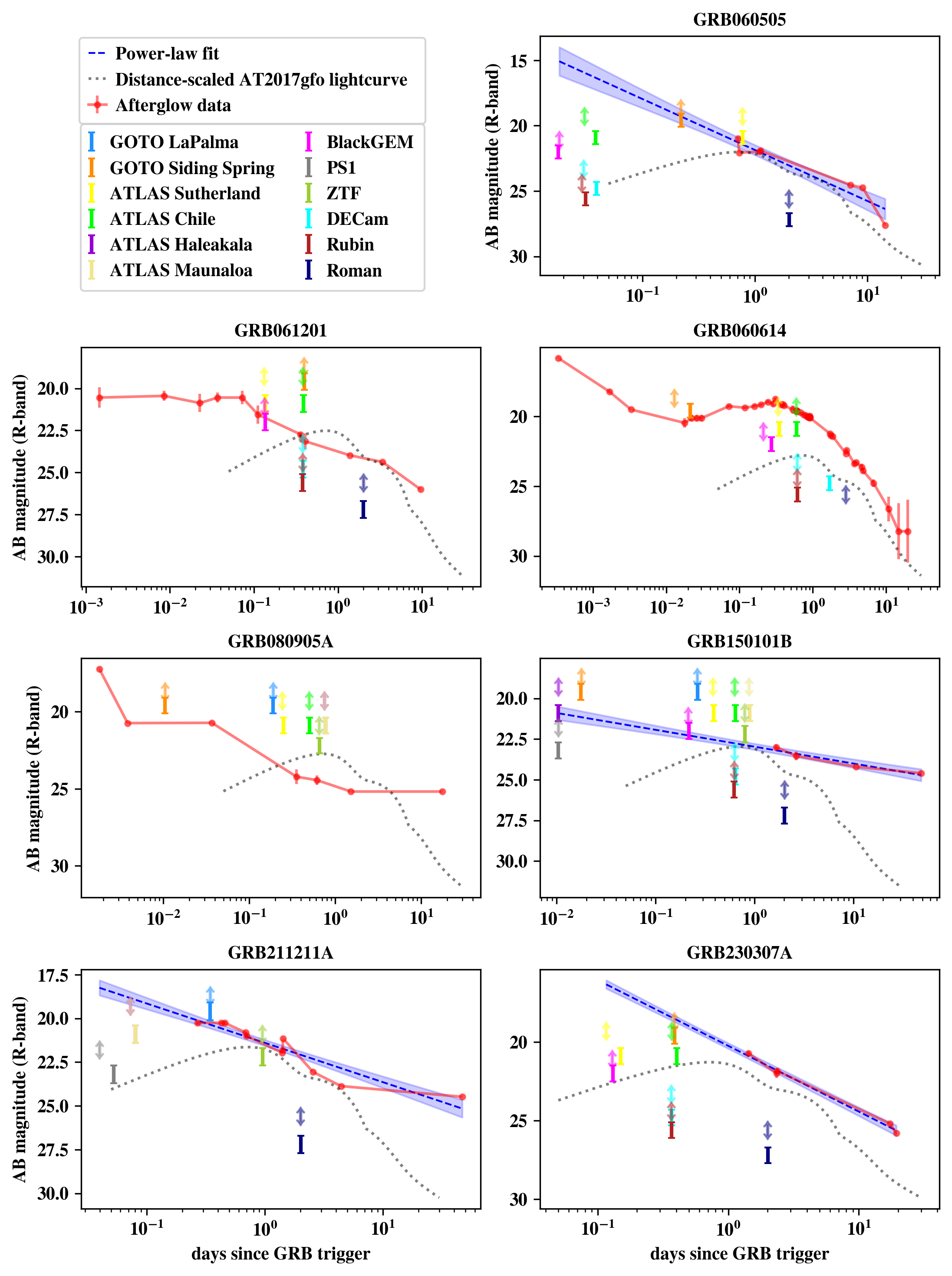}
    \caption{NSBH Detectability: The afterglow lightcurves of GRBs with double-arrow (rook-shaped) markers representing the time of source-tile observation and the limiting magnitude of telescopes for shallow (deep) searches in an O5 scenario with NSBH progenitor. The markers are color-coded to represent different telescopes, as shown in the plot legend. The shaded blue regions show the uncertainty associated with power-law extrapolation of the afterglow lightcurves.}
    \label{fig:NSBH_o5_detectability}
\end{figure*}

Figures \ref{fig:BNS_o5_detectability} and \ref{fig:NSBH_o5_detectability} show the limiting magnitudes and time delays between the event trigger time and source tile observation for different telescopes that successfully observe the source tile within a total observing time of 50 hours, assuming tiling of the 90\% credible region. Note that the detections reported below also assume that EM searches would have been carried out for all events in our sample without imposing any SNR-based GW detection cutoffs. 

In the BNS case, both the afterglows and kilonovae of all 7 events are detected by at least one telescope in our list (see Figure \ref{fig:BNS_o5_detectability}).
On the other hand, in the NSBH case, counterparts are detected in only 6 out of 7 events, with the exception being GRB 080905A (as seen in Figure \ref{fig:NSBH_o5_detectability}). Here, the deep telescopes capable of detecting its counterparts do not observe the source tile owing to a combination of fixed-grid tiling and the 90\% area cutoff (see Section~\ref{sec: discussions-tiling-problems} for details). Overall, telescopes with deeper limiting magnitudes like DECam, Rubin, and Roman seem to play a crucial role in all the recorded detections given their depth, as the afterglow and kilonova brightness fade to average magnitudes of $22.19 \pm 1.75$ and $22.41 \pm 0.64$, respectively, by day 1. 

It is worth noting that the source tile is successfully captured even for events with very large localisation areas, such as GRB 150101B in the BNS case (292 deg$^2$), and GRB 060614 in the BNS and NSBH cases (23554 deg$^2$ and 591 deg$^2$ respectively). This is because the source location generally lies within the first few tiles, as expected from the probability-based ranked tiling strategy. For example, for GRB 060614 in the NSBH scenario (591 deg$^2$), the true position falls within tile rank 4 out of 21 for the wide-field surveys GOTO and ATLAS, and within the top few tens of ranked tiles for smaller-FOV facilities such as Rubin, Pan-STARRS, DECam, BlackGEM, and Roman, despite these requiring hundreds to thousands of tiles to cover the full 591 deg$^2$ region. As another example, GRB 080905A's source location in the BNS scenario (20 deg$^2$) lies within the first tile for all three small FOV facilities DECam, BlackGEM, and Roman, which would need 10, 11, and 75 tiles, respectively, to cover the 90\% credible region. 

 Across all skymaps, the delay between shallow and deep exposures (see Table \ref{tab:telescopes}) in capturing the source tile is minimal, usually of order 5–15 minutes, although a few outliers extend to 1–2 days. In extreme cases, for large skymaps attempted with small–FOV facilities, the deep exposure search fails to schedule the source tile within the 50-hour observing window even when it has a relatively high rank in the probability-ranked tile list. Overall, the increase in detections between shallow and deep exposure searches is modest. Only a subset of skymaps benefits from the deeper exposures, primarily where the greater depth allows an otherwise marginally detectable counterpart to be confidently detected. This trade-off between depth and sky coverage is reflected in the detection statistics below.

For afterglows, the total number of detections increases from 59 to 64 across all telescopes and events. In the BNS scenario, additional afterglow detections occur for GOTO Siding Spring (GRB~060505), BlackGEM (GRB~061201), PS1 (GRB~150101B), and DECam (GRB~080905A), while DECam (GRB~060614) and Roman (GRB~150101B) each lose one detection. In the NSBH scenario, additional afterglow detections are seen for GOTO Siding Spring (GRB~060505), ATLAS Haleakala (GRB~150101B), and BlackGEM (GRB~061201 and GRB~150101B), while Roman (GRB~060614) loses one detection. For kilonovae, the number of detections increases from 29 to 30. In the BNS scenario, deeper exposures benefit PS1 (GRB~080905A) and ZTF (GRB~211211A), while Roman (GRB~150101B) loses one detection. In the NSBH scenario, deeper exposures improve detection only for ZTF (GRB~211211A), whereas Roman (GRB~060614) loses one detection.

More than 73\% of detections, spanning all telescopes, progenitor types (BNS and NSBH), and counterpart classes (afterglow and kilonova), occur within the first 24 hours of the GW trigger. Across all BNS and NSBH cases, afterglow detections are concentrated early. In shallow exposures, out of 59 detections, 15 (25\%) occur within 0.1 days, 32 (54\%) occur between 0.1 and 1 day, and 12 (20\%) occur beyond 1 day. In deep exposures, out of 64 detections, the corresponding counts are 15 (23\%), 36 (56\%), and 13 (20\%). In contrast, kilonova detections are delayed. In shallow exposures, out of 29 detections, 17 (59\%) occur between 0.1 and 1 day, and 12 (41\%) occur beyond 1 day. In deep exposures, out of 30 detections, the corresponding counts are 18 (60\%) and 12 (40\%). No kilonova detections are recorded within 0.1 days of the trigger, consistent with their slower rise and intrinsically fainter emission compared to afterglows.

\section{Discussions} \label{sec: discussions}
\subsection{Merger-induced GRBs during O5}

The simulated GW skymaps of merger-induced GRBs in our sample have relatively small localisations in O5. More than half of all events have network SNR $>12$, whereas the remainder lie in the subthreshold regime (see Section~\ref{sec: results-skymaps}), indicating that they are still detectable by the GW network rather than being entirely undetectable. The median localisation areas are 20 deg$^2$ for the BNS scenario and 5 deg$^2$ for the NSBH scenario (Table~\ref{tab:gw-statistics}), with several events localised to only a few square degrees. These values are comparable to, or even smaller than, the localisation achieved for GW170817. We emphasize that these results apply specifically to on-axis, EM-bright BNS and NSBH mergers rather than to the broader compact binary merger population expected during O5. They also correspond to an optimistic configuration of the four-detector LIGO–Virgo–KAGRA network operating near the higher end of its projected O5 sensitivity. Population-level forecasts typically quote median localisation areas of hundreds to thousands of square degrees for BNS and NSBH mergers \citep{
abbott2020Prospects, petrov2022Datadriven}, reflecting the full population of mergers, including many distant and unfavourably oriented systems.

To place our sample in a broader context, we compare the observed GRB rate to expectations based on the LVK population measurements \citep{abac2025GWTC40}. Over the past $\sim$22 years, seven GRBs with secure redshifts $z<0.15$ and plausible merger origins have been detected. Correcting for the sky coverage of the gamma-ray instruments over this period yields an effective all-sky exposure of $\sim$13.3 years (assuming $\sim$1/6 sky coverage for \textit{Swift} during the first 4 years and $\sim$70\% coverage for \textit{Fermi}/GBM thereafter\footnote{based on \citealt{barthelmy2005Burst} and \citealt{meegan2009Fermi}}). This corresponds to an observed on-axis GRB rate of $\sim$0.53 year$^{-1}$. For a sample of seven events, the 90\% Poisson interval corresponds to $\sim$0.25$-$0.99 year$^{-1}$. However, we also note that redshift completeness, in particular for {\em Fermi} is very poor, such that the numbers represent hard lower limits. The latest LVK population analysis finds a BNS merger-rate density of $\mathcal{R}_{\rm BNS}=49^{+121}_{-42}$ Gpc$^{-3}$ year$^{-1}$ under the Binned Gaussian Process model \citep{abac2025GWTC40}. Within a sphere of radius 635 Mpc (corresponding to the largest GRB distance in our sample), this corresponds to an intrinsic BNS merger rate of $\sim53^{+130}_{-45}$ year$^{-1}$. However, only a fraction of these mergers are expected to produce observable GRBs. Following \citet{sarin2022Linking}, the fraction of BNS mergers that produce successful GRB jets is estimated to be $f_{\rm jet,BNS}=0.69^{+0.27}_{-0.37}$, while the inferred jet opening angle of $\theta_j \approx 15^{\circ}$ corresponds to an on-axis beaming fraction $f_{\rm beam}\approx 0.034$. Applying these corrections implies an expected observable GRB rate of order $\sim$1.2 year$^{-1}$ within this volume.

Under the simplifying assumption that our GRB sample is dominated by BNS mergers, this is broadly consistent with the rate inferred from our GRB sample, given the large statistical and systematic uncertainties. Nevertheless, our sample may contain both BNS and NSBH progenitors. Population-synthesis models suggest that the fraction of NSBH mergers that successfully launch jets may be significantly lower ($\sim$0.5–10\%; \citealt{sarin2022Linking, broekgaarden2021Impact}), implying that only a small fraction of NSBH mergers produce observable GRBs. However, this fraction may be higher for systems with lower black hole masses or more comparable mass ratios, such as GW230529 \citep{abac2024Observation}. If some fraction of the observed GRB sample indeed arises from NSBH mergers, the correspondence between the GRB rate inferred above and the BNS-only expectation would therefore be less direct, introducing additional uncertainty in interpreting the observed GRB rate in terms of the underlying compact binary merger population.

GRB observations trace only the subset of mergers that successfully launch relativistic jets toward the observer. Mergers with similar viewing geometries that fail to produce observable GRBs would still generate comparable GW signals and localisation properties. Taken together, these comparisons suggest that GRB-like compact object mergers represent a small but astrophysically meaningful subset of the overall compact binary merger population, and that a fraction of favourably oriented mergers during O5 may therefore produce comparatively small GW skymaps similar to those explored in this work.

\subsection{Follow-up of optical counterparts}

In our simulations, the GW localisation regions are small in most cases, and both the afterglow and kilonova counterparts are detected by at least one telescope in the follow-up simulation, even though the counterparts are distant and intrinsically faint (see Section~\ref{sec: results-detectability}). Because the source tile typically lies within the highest-probability region of the skymap and therefore has a high rank, it is usually observed early in the follow-up sequence, even when the overall localisation area is large. As a result, the primary challenges for follow-up shift from sky coverage to the depth of observations and the reliable identification of the correct transient among many candidates.

\subsubsection{Deep telescopes are crucial}\label{sec: discussions-deep-telescopes}

A closer examination of Figures \ref{fig:BNS_o5_detectability} and \ref{fig:NSBH_o5_detectability} shows which telescopes successfully recover the faint counterparts in our sample. While several shallower facilities detect the brighter afterglows in the early hours, the deeper facilities more consistently recover both the afterglow and the kilonova. In our sample, 3 out of 7 afterglows are already fainter than 22 AB magnitude within 0.1 days (2.4 hours). The afterglow and kilonova brightness fade to average magnitudes of $22.19 \pm 1.75$ and $22.41 \pm 0.64$, respectively, by day 1. Therefore, telescopes capable of reaching depths of $\gtrsim 23$ mag become important as counterparts fade further. In particular, DECam, Rubin, and Roman contribute a large fraction of the successful detections due to their significantly deeper limiting magnitudes compared to shallower facilities.

As reported in Section~\ref{sec: results-detectability}, the overall increase in recorded detections from deeper exposures is modest, with afterglow detections increasing from 59 to 64 and kilonova detections from 29 to 30. The detection improvement is slightly more pronounced for afterglows than for kilonovae because kilonovae are fainter. Many telescopes we consider remain too shallow to detect the faintest counterparts even in their deeper configuration, while the telescopes that already reach sufficient depth gain relatively little by going deeper. This reinforces that the key distinction is therefore not simply between shallow and deep exposures, but between facilities that can or cannot reach a limiting magnitude of roughly 23 mag. Once this depth is achieved, both afterglow and kilonova counterparts are frequently detectable, and below it, they are often missed.

Furthermore, our kilonova detectability estimates based on the distance-scaled lightcurve of AT2017gfo (see Section~\ref{subsec: kilonova-template}) do not capture the full diversity of kilonovae. AT2017gfo is intrinsically brighter than some kilonova candidates associated with GRBs \citep[e.g.,][]{gompertz2018Diversity}. Our detectability estimates may therefore be somewhat optimistic, as fainter kilonovae would be more challenging to detect, further emphasising the importance of deep telescopes.

\subsubsection{Identifying counterparts is challenging}\label{sec: discussions-discoverability}

While the counterparts in our simulations are detectable by one or more telescopes, correctly identifying a given transient as the GW counterpart remains a non-trivial exercise. The nearby distance and small localisation of GW170817 resulted in a small number of expected transients within the 3D localisation volume, but this is not, in general, the case. In O5, higher source distances, sometimes in combination with larger error boxes, imply that many unrelated transients are expected within a given localisation region. Furthermore, identifying the transient with a given host galaxy may be non-trivial. For example, within our GRB sample at $z<0.15$, only one of the putative host galaxies had a redshift prior to the GRB detection (GRB~061201). However, for events with localisations comparable to those in our simulations, multiple revisits of the same fields should be possible, improving the prospects for identifying the correct transient as it evolves.

\begin{figure}
\includegraphics[width=\columnwidth]{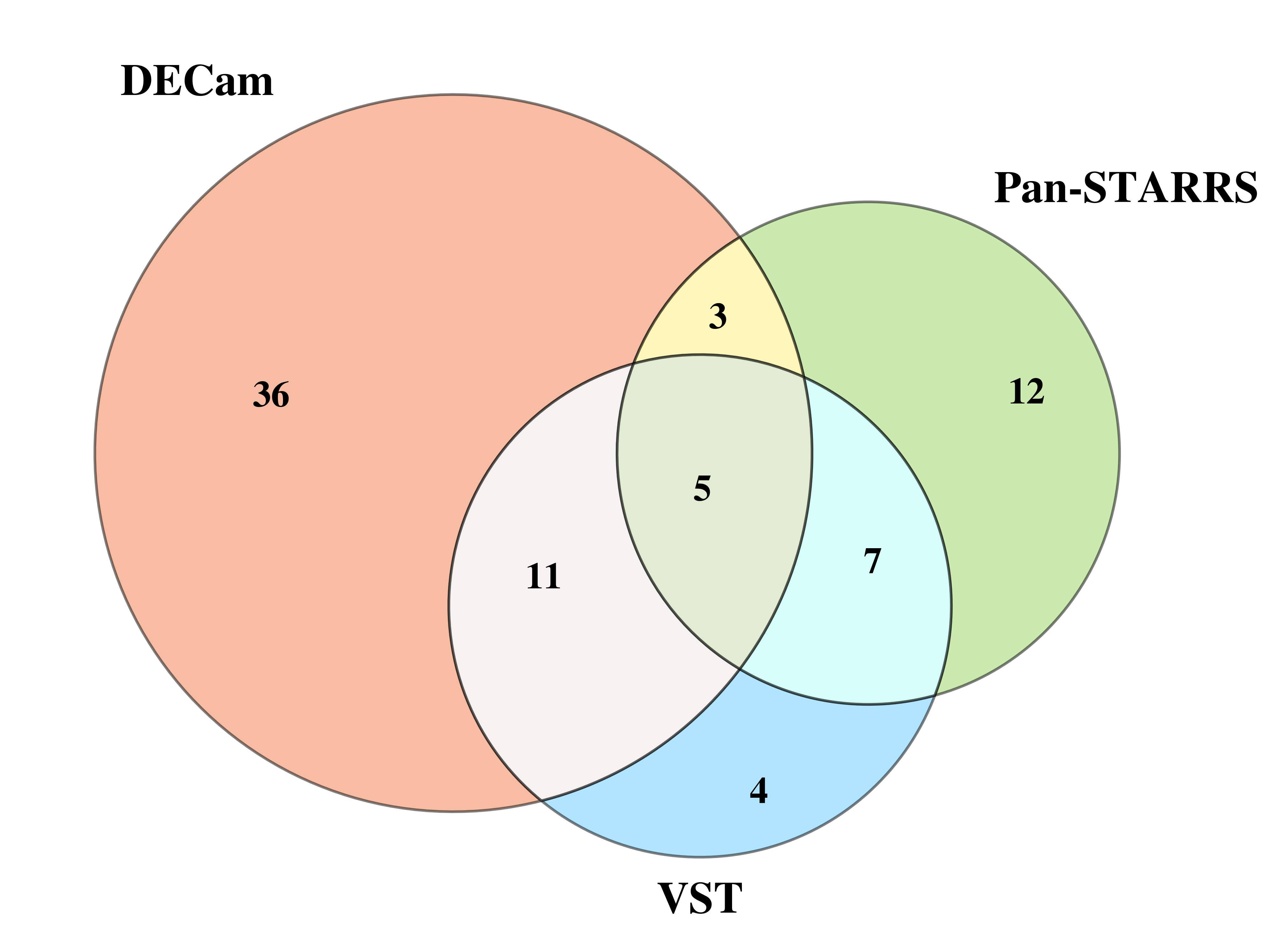}
    \caption{Overlap of transients identified within the GW190814 localisation by Pan-STARRS, VST, and DECam surveys. The small intersection demonstrates that different surveys recover different subsets of transients within the same localisation region.}
    \label{fig:gw190814_venn}
\end{figure}

A useful comparison for the challenges of identifying the correct counterpart can be found in searches of GW error boxes to date. A particularly illustrative example is the follow-up of GW190814 \citep{abbott2020GW190814}, an event localised to 18.5~deg$^2$. This object was the result of the merger of two compact objects with masses of 23.2 and 2.6 M$_{\odot}$ and, as such, was not expected to drive luminous electromagnetic emission \citep{foucart2018Remnant}. Follow-up searches were undertaken by several groups, including VLT Survey Telescope (VST; \citealt{capaccioli2011VLT}), ZTF, Pan-STARRS, GOTO, ATLAS, and DECam. These surveys identified a large number of transient sources within the error box \citep[e.g.,][]{ackley2020Observational}. However, while all surveys searched the same region of sky, not all surveys identified the same transients, consistent with expectations of non-identical transient recovery in GW follow-up campaigns \citep{nissanke2013Identifying}. In Figure~\ref{fig:gw190814_venn}, we show a Venn diagram of the transients identified by Pan-STARRS, VST, and DECam. Only 5 transients were reported by all three surveys, demonstrating that the recovery fraction of transients is far from 100\% for any of the surveys considered, although it should be noted that some transients may be variable stars that are more effectively removed by some searches than others. It is also possible that some transients were discovered but not publicly reported in the General Coordinates Network (GCN) or AstroNotes. Without a detailed analysis of the pixel data, which is beyond the scope of this paper, we cannot infer the reason for the small overlap in Fig.~\ref{fig:gw190814_venn}. However, factors such as chip gaps, poor quality image subtractions in the centres of bright host galaxies, and erroneous flagging of marginal detections by real/bogus classifiers may all contribute.

\subsection{Insights for follow-up campaigns}

\subsubsection{Non-linear probability--area structure of GW skymaps}

A key insight for the follow-up of GW skymaps comes from the probability–area curves shown in the bottom-right panels of Figures \ref{fig:BNS-O5-skymaps} and \ref{fig:NSBH-O5-skymaps}. These curves reveal a strong non-linear relationship between enclosed probability and credible region area that is not widely emphasised in the literature. For example, a 90\% region spanning 1800 deg$^2$ does not imply that the 75\% region spans $\sim$1500 deg$^2$. In some cases, it may instead be as small as 200 deg$^2$. 

Thus, even for events with larger localisations, the 50–75\% probability core remains comparatively compact and often within reach for many telescopes. This insight is particularly relevant for facilities with small FOVs, such as Roman, which could target the central high-probability region even for events with large 90\% localisation areas, rather than restricting ToO triggers to events with small skymaps. Follow-up strategies should therefore prioritise the highest-probability tiles, which typically form a compact region of the skymap, as this provides the most efficient observational strategy compared to covering spatially extended regions that merely sum to the same cumulative probability.

\subsubsection{Limitations of fixed probability cutoffs in tiling}\label{sec: discussions-tiling-problems}

As reported in Section~\ref{sec: results-detectability}, applying a strict 90\% cutoff in combination with a fixed-grid tiling scheme and an automated scheduler can result in the source tile not being observed when the true source lies near the boundary of the 90\% credible region. In a fixed-grid approach, the sky is partitioned into a predefined set of tiles, and tiles are selected in probability rank order until the cumulative probability of the selected tiles reaches 90\%. If the true source location lies in a region covered only by tiles beyond this cutoff, the source tile is simply not observed.

This behaviour is evident in the NSBH skymap for GRB~080905A (90\% area 5~deg$^2$), where the true source lies at $\sim$97\% enclosed probability and is therefore excluded from the tiles selected to reach the 90\% threshold. As a result, deep facilities with small FOVs that are capable of detecting the counterpart, such as Roman and DECam, do not schedule the source tile despite the counterpart being bright enough to detect. A similar effect occurs for DECam in the BNS case of GRB~230307A (90\% area of 2~deg$^2$), where the true source lies at the 91\% credible level. Although DECam has a 3~deg$^2$ FOV, the source tile (rank 2 in the probability-ordered list) falls just beyond the strict 90\% cutoff and is therefore excluded from the selected tiles.

For events with small localisations, extending coverage slightly beyond the nominal probability threshold provides a simple mitigation. In such cases, covering the 99\% region, typically only a factor of two to three larger than the 90\% region, offers a more robust strategy against these tiling edge effects.

\subsubsection{Rapid response is critical for telescopes reaching $\lesssim22$ mag}  

As discussed in Section~\ref{sec: discussions-deep-telescopes}, deep observations are essential for recovering the faint afterglow and kilonova emission at the distances considered here. This naturally raises the question of where shallower surveys still have an opportunity. Facilities such as ZTF, ATLAS, and GOTO typically reach limiting magnitudes of $\sim$19--22 mag and are therefore at a disadvantage when searching for kilonovae or afterglows at later epochs for the events we consider. 

Their primary opportunity lies in detecting the bright early afterglow within the first few hours after the GW trigger. This is also evident in Figures~\ref{fig:BNS_o5_detectability} and \ref{fig:NSBH_o5_detectability}, where most detections by telescopes with limiting magnitude $\lesssim22$ mag occur at early times. Our simulations assume an optimistic latency of 15 minutes between the GW trigger and the start of observations. In practice, response times are often of the order of hours, depending on source visibility and scheduling constraints, meaning that some of the early afterglow detections seen in our simulations may be optimistic for such facilities. Nevertheless, these detections underscore the need for a rapid response for capturing the early afterglow emission.

\subsubsection{Limitations of SNR and distance-based prioritisation}

 As shown in Section~\ref{sec: results-skymaps}, several events deviate significantly from the typical SNR and distance-based scaling expectations. Relations such as sky localisation $\propto 1/{\rm SNR}^2$ or ${\rm SNR}\propto 1/{\rm D_L}$ provide only approximate guidance and do not reliably capture event-to-event variations in network geometry and sky position. The counterexamples in Table~\ref{tab:gw-statistics} illustrate that events with similar distances or SNRs can nonetheless have localisation areas differing by orders of magnitude.  Consequently, prioritisation strategies for follow-up should not discount any events solely on the basis of distance or network SNR.

\section{Conclusions}\label{sec: conclusions}

In this work, we investigated the prospects for detecting optical counterparts to compact object mergers that produce GRBs during the upcoming observing run O5. We compiled a sample of nearby GRBs ($z<0.15$) with plausible merger origins and simulated how these events would appear to the LVK detector network operating at the higher end of its projected O5 sensitivity. Using the resulting GW skymaps, we then simulated optical follow-up with a range of existing and upcoming facilities and assessed whether the associated afterglow and kilonova emission would be detectable.

Our simulations show that such events can be localised to relatively small regions of the sky. The median localisation areas are 20 deg$^2$ for the BNS scenario and 5 deg$^2$ for the NSBH scenario. The source location typically lies within the highest-probability region of the skymap and is therefore observed early in the follow-up sequence, even for events with larger localisation areas. As a result, the optical counterparts in our sample are often detectable. Both afterglows and kilonovae are recovered by at least one telescope in nearly all cases. However, the results show that detectability is strongly influenced by observational depth. By day 1 after the trigger, the afterglows and kilonovae in our sample are typically fainter than 22 mag, making facilities capable of reaching depths of $\gtrsim$ 23 mag particularly important for recovering faint counterparts. In our simulations, deeper facilities such as DECam, Rubin, and Roman contribute a substantial fraction of the successful detections, while shallower surveys primarily recover only the brighter and earlier afterglow emission.

While detection itself appears feasible in most cases, identifying the correct transient within the localisation region remains challenging. Even for comparatively small GW skymaps, multiple unrelated transients may be present, and different surveys may recover different subsets of these sources. Securely identifying the true counterpart, therefore, requires repeated observations and careful vetting of candidates based on their temporal evolution and host-galaxy associations.

Finally, our simulations highlight several practical considerations for follow-up campaigns. The probability distribution within GW skymaps is strongly non-linear, meaning that a large fraction of the localisation probability is often concentrated within a comparatively compact core. Prioritising observations of these highest-probability regions provides an efficient strategy for follow-up, even when the total 90\% localisation area is larger. In addition, strict probability cutoffs in fixed-grid tiling strategies can occasionally exclude the true source position, suggesting that modest extensions beyond the nominal probability threshold may improve recovery.

These results suggest that compact object mergers producing observable GRBs during O5 may yield GW localisation regions that are substantially smaller than those expected for the broader merger population, and that their optical counterparts can often be detected with available follow-up facilities. In such cases, the primary challenge for multimessenger discovery lies less in GW localisation itself and more in achieving sufficient observational depth and implementing effective strategies for identifying the true counterpart among unrelated transients.

\section*{Acknowledgements}
 K.K. thanks Aditya Vijaykumar, Nikhil Sarin, and Uddipta Bhardwaj for insightful discussions on Bilby that helped resolve skymap-related issues. K.K. acknowledges support from the Radboud Scholarship Programme and the Radboud Honours Academy Beyond the Frontiers Grant. K.K. and S.N. acknowledge support from Deutsches Elektronen-Synchrotron (DESY). A.J.L. was supported by the European Research Council (ERC) under the European Union’s Horizon 2020 research and innovation programme (grant agreement No.~725246). M.F. acknowledges financial support of Taighde \'{E}ireann – Research Ireland under Grant number 24/FFP-P/12959. S.~A. is supported by an LSST-DA Catalyst Fellowship; this publication was thus made possible through the support of Grant 62192 from the John Templeton Foundation to LSST-DA. S.~A. also gratefully acknowledges support from Stanford University, the United States Department of Energy, and a generous grant from Fred Kavli and The Kavli Foundation. I.A. is supported by the National Science Foundation award AST 2505775, NASA grant 24-ADAP24-0159, Scialog award SA-LSST-2024-102a, and the Discovery Alliance Catalyst Fellowship Mentors award 2025-62192-CM-19. A.G. acknowledges the financial support from the Slovenian Research Agency (grants P1-0031, I0-0033, J1-2460, N1-0344) and the SNN-147362 OTKA grant of the Hungarian Research, Development and Innovation Office. P.S. acknowledges support from STFC grants ST/V005677/1 and ST/Y00423X/1. This work was initially conceived at The Unconventional Thinking Tank Conference 2022, which is supported by INAF.

%%%%%%%%%%%%%%%%%%%%%%%%%%%%%%%%%%%%%%%%%%%%%%%%%%
\section*{Data Availability}

Public data release for this paper including afterglow lightcurves, skymaps from GW simulations, and scripts for using the scheduler and making figures is on Zenodo: \url{https://doi.org/10.5281/zenodo.10892780}. The code used for tiling and scheduling observations, which is a modified version of Shaon Ghosh's code, is available on the git repository: \url{https://github.com/krishnakruthi/sky_tiling}.



\bibliographystyle{mnras}
\bibliography{msc_thesis} % if your bibtex file is called example.bib

@inproceedings{bloemen2015BlackGEM,
	address = {San Francisco},
	series = {Astronomical {Society} of the {Pacific} {Conference} {Series}},
	title = {The {BlackGEM} {Array}: {Searching} for {Gravitational} {Wave} {Source} {Counterparts} to {Study} {Ultra}-{Compact} {Binaries}},
	volume = {496},
	shorttitle = {The {BlackGEM} {Array}},
	url = {https://ui.adsabs.harvard.edu/abs/2015ASPC..496..254B},
	number = {496},
	urldate = {2023-12-28},
	booktitle = {Living {Together}: {Planets}, {Host} {Stars} and {Binaries}},
	author = {Bloemen, S. and Groot, P. and Nelemans, G. and Klein-Wolt, M.},
	month = jul,
	year = {2015},
	note = {Conference Name: Living Together: Planets, Host Stars and Binaries
ADS Bibcode: 2015ASPC..496..254B},
	pages = {254},
}

@article{petrov2022Datadriven,
	title = {Data-driven {Expectations} for {Electromagnetic} {Counterpart} {Searches} {Based} on {LIGO}/{Virgo} {Public} {Alerts}},
	volume = {924},
	url = {http://arxiv.org/abs/2108.07277},
	doi = {10.3847/1538-4357/ac366d},
	number = {2},
	urldate = {2022-12-05},
	journal = {The Astrophysical Journal},
	publisher = {American Astronomical Society},
	author = {Petrov, Polina and Singer, Leo P. and Coughlin, Michael W. and Kumar, Vishwesh and Almualla, Mouza and Anand, Shreya and Bulla, Mattia and Dietrich, Tim and Foucart, Francois and Guessoum, Nidhal},
	month = jan,
	year = {2022},
	note = {136 citations (INSPIRE 2026/4/1)
100 citations w/o self (INSPIRE 2026/4/1)
arXiv:2108.07277 [astro-ph.HE]},
	pages = {54},
}

@article{nissanke2013Identifying,
	title = {Identifying {Elusive} {Electromagnetic} {Counterparts} to {Gravitational} {Wave} {Mergers}: an end-to-end simulation},
	volume = {767},
	url = {http://arxiv.org/abs/1210.6362},
	doi = {10.1088/0004-637X/767/2/124},
	number = {2},
	urldate = {2022-12-05},
	journal = {Astrophysical Journal},
	publisher = {Institute of Physics Publishing},
	author = {Nissanke, Samaya and Kasliwal, Mansi and Georgieva, Alexandra},
	year = {2013},
	note = {218 citations (INSPIRE 2026/4/1)
183 citations w/o self (INSPIRE 2026/4/1)
arXiv:1210.6362 [astro-ph.HE]},
	pages = {124},
}

@article{ghosh2016Tiling,
	title = {Tiling strategies for optical follow-up of gravitational-wave triggers by telescopes with a wide field of view},
	volume = {592},
	url = {http://arxiv.org/abs/1511.02673},
	doi = {10.1051/0004-6361/201527712},
	urldate = {2022-12-05},
	journal = {Astronomy and Astrophysics},
	publisher = {EDP Sciences},
	author = {Ghosh, Shaon and Bloemen, Steven and Nelemans, Gijs and Groot, Paul J. and Price, Larry R.},
	month = aug,
	year = {2016},
	note = {44 citations (INSPIRE 2026/4/1)
37 citations w/o self (INSPIRE 2026/4/1)
arXiv:1511.02673 [astro-ph.IM]},
	pages = {A82},
}

@article{margutti2021First,
	title = {First {Multimessenger} {Observations} of a {Neutron} {Star} {Merger}},
	volume = {59},
	url = {http://arxiv.org/abs/2012.04810},
	doi = {10.1146/annurev-astro-112420-030742},
	urldate = {2022-12-02},
	journal = {Annual Review of Astronomy and Astrophysics},
	publisher = {Annual Reviews Inc.},
	author = {Margutti, Raffaella and Chornock, Ryan},
	month = jun,
	year = {2021},
	note = {154 citations (INSPIRE 2026/4/1)
136 citations w/o self (INSPIRE 2026/4/1)
arXiv:2012.04810 [astro-ph.HE]},
	pages = {155--202},
}

@article{metzger2010Electromagnetic,
	title = {Electromagnetic {Counterparts} of {Compact} {Object} {Mergers} {Powered} by the {Radioactive} {Decay} of {R}-process {Nuclei}},
	volume = {406},
	issn = {00358711},
	url = {http://arxiv.org/abs/1001.5029},
	doi = {10.1111/j.1365-2966.2010.16864.x},
	number = {4},
	urldate = {2026-01-12},
	journal = {Monthly Notices of the Royal Astronomical Society},
	author = {Metzger, B.D. and Martinez-Pinedo, G. and Darbha, S. and Quataert, E. and Arcones, A. and Kasen, D. and Thomas, R. and Nugent, P. and Panov, I.V. and Zinner, N.T.},
	year = {2010},
	note = {1099 citations (INSPIRE 2026/4/1)
946 citations w/o self (INSPIRE 2026/4/1)
arXiv:1001.5029 [astro-ph.HE]},
	pages = {2650},
}

@article{troja2019Afterglow,
	title = {The afterglow and kilonova of the short {GRB} {160821B}},
	volume = {489},
	issn = {0035-8711, 1365-2966},
	url = {http://arxiv.org/abs/1905.01290},
	doi = {10.1093/mnras/stz2255},
	number = {2},
	urldate = {2023-08-04},
	journal = {Monthly Notices of the Royal Astronomical Society},
	author = {Troja, E. and Castro-Tirado, A.J. and Becerra Gonzalez, J. and {others}},
	month = oct,
	year = {2019},
	note = {160 citations (INSPIRE 2026/4/1)
133 citations w/o self (INSPIRE 2026/4/1)
arXiv:1905.01290 [astro-ph.HE]},
	pages = {2104--2116},
}

@article{singer2016Rapid,
	title = {Rapid {Bayesian} position reconstruction for gravitational-wave transients},
	volume = {93},
	url = {http://arxiv.org/abs/1508.03634},
	doi = {10.1103/PhysRevD.93.024013},
	number = {2},
	urldate = {2022-12-05},
	journal = {Physical Review D},
	publisher = {American Physical Society},
	author = {Singer, Leo P. and Price, Larry R.},
	month = jan,
	year = {2016},
	note = {450 citations (INSPIRE 2026/4/1)
389 citations w/o self (INSPIRE 2026/4/1)
arXiv:1508.03634 [gr-qc]},
	pages = {024013},
}

@article{levan2024Heavyelement,
	title = {Heavy-element production in a compact object merger observed by {JWST}},
	volume = {626},
	copyright = {2023 The Author(s)},
	issn = {1476-4687},
	url = {https://www.nature.com/articles/s41586-023-06759-1},
	doi = {10.1038/s41586-023-06759-1},
	language = {en},
	number = {8000},
	urldate = {2026-04-01},
	journal = {Nature},
	publisher = {Nature Publishing Group},
	author = {Levan, Andrew J. and Gompertz, Benjamin P. and Salafia, Om Sharan and Bulla, Mattia and Burns, Eric and Hotokezaka, Kenta and Izzo, Luca and Lamb, Gavin P. and Malesani, Daniele B. and Oates, Samantha R. and Ravasio, Maria Edvige and Rouco Escorial, Alicia and Schneider, Benjamin and Sarin, Nikhil and Schulze, Steve and Tanvir, Nial R. and Ackley, Kendall and Anderson, Gemma and Brammer, Gabriel B. and Christensen, Lise and Dhillon, Vikram S. and Evans, Phil A. and Fausnaugh, Michael and Fong, Wen-fai and Fruchter, Andrew S. and Fryer, Chris and Fynbo, Johan P. U. and Gaspari, Nicola and Heintz, Kasper E. and Hjorth, Jens and Kennea, Jamie A. and Kennedy, Mark R. and Laskar, Tanmoy and Leloudas, Giorgos and Mandel, Ilya and Martin-Carrillo, Antonio and Metzger, Brian D. and Nicholl, Matt and Nugent, Anya and Palmerio, Jesse T. and Pugliese, Giovanna and Rastinejad, Jillian and Rhodes, Lauren and Rossi, Andrea and Saccardi, Andrea and Smartt, Stephen J. and Stevance, Heloise F. and Tohuvavohu, Aaron and van der Horst, Alexander and Vergani, Susanna D. and Watson, Darach and Barclay, Thomas and Bhirombhakdi, Kornpob and Breedt, Elmé and Breeveld, Alice A. and Brown, Alexander J. and Campana, Sergio and Chrimes, Ashley A. and D’Avanzo, Paolo and D’Elia, Valerio and De Pasquale, Massimiliano and Dyer, Martin J. and Galloway, Duncan K. and Garbutt, James A. and Green, Matthew J. and Hartmann, Dieter H. and Jakobsson, Páll and Kerry, Paul and Kouveliotou, Chryssa and Langeroodi, Danial and Le Floc’h, Emeric and Leung, James K. and Littlefair, Stuart P. and Munday, James and O’Brien, Paul and Parsons, Steven G. and Pelisoli, Ingrid and Sahman, David I. and Salvaterra, Ruben and Sbarufatti, Boris and Steeghs, Danny and Tagliaferri, Gianpiero and Thöne, Christina C. and de Ugarte Postigo, Antonio and Kann, David Alexander},
	month = feb,
	year = {2024},
	pages = {737--741},
}

@article{hinderer2010Tidal,
	title = {Tidal deformability of neutron stars with realistic equations of state and their gravitational wave signatures in binary inspiral},
	volume = {81},
	issn = {1550-7998, 1550-2368},
	url = {http://arxiv.org/abs/0911.3535},
	doi = {10.1103/PhysRevD.81.123016},
	number = {12},
	urldate = {2026-03-25},
	journal = {Physical Review D},
	author = {Hinderer, Tanja and Lackey, Benjamin D. and Lang, Ryan N. and Read, Jocelyn S.},
	month = jun,
	year = {2010},
	note = {arXiv:0911.3535 [astro-ph]},
	pages = {123016},
}

@article{abac2024Observation,
	title = {Observation of {Gravitational} {Waves} from the {Coalescence} of a 2.5–4.5 {M}⊙ {Compact} {Object} and a {Neutron} {Star}},
	volume = {970},
	issn = {2041-8205},
	url = {https://doi.org/10.3847/2041-8213/ad5beb},
	doi = {10.3847/2041-8213/ad5beb},
	language = {en},
	number = {2},
	urldate = {2026-03-25},
	journal = {The Astrophysical Journal Letters},
	publisher = {The American Astronomical Society},
	author = {Abac, A. G. and Abbott, R. and Abouelfettouh, I. and Acernese, F. and Ackley, K. and Adhicary, S. and Adhikari, N. and Adhikari, R. X. and Adkins, V. K. and Agarwal, D. and Agathos, M. and Abchouyeh, M. Aghaei and Aguiar, O. D. and Aguilar, I. and Aiello, L. and Ain, A. and Ajith, P. and Akçay, S. and Akutsu, T. and Albanesi, S. and Alfaidi, R. A. and Al-Jodah, A. and Alléné, C. and Allocca, A. and Al-Shammari, S. and Altin, P. A. and Alvarez-Lopez, S. and Amato, A. and Amez-Droz, L. and Amorosi, A. and Amra, C. and Ananyeva, A. and Anderson, S. B. and Anderson, W. G. and Andia, M. and Ando, M. and Andrade, T. and Andres, N. and Andrés-Carcasona, M. and Andrić, T. and Anglin, J. and Ansoldi, S. and Antelis, J. M. and Antier, S. and Aoumi, M. and Appavuravther, E. Z. and Appert, S. and Apple, S. K. and Arai, K. and Araya, A. and Araya, M. C. and Areeda, J. S. and Argianas, L. and Aritomi, N. and Armato, F. and Arnaud, N. and Arogeti, M. and Aronson, S. M. and Arun, K. G. and Ashton, G. and Aso, Y. and Assiduo, M. and de Souza Melo, S. Assis and Aston, S. M. and Astone, P. and Attadio, F. and Aubin, F. and AultONeal, K. and Avallone, G. and Azrad, D. and Babak, S. and Badaracco, F. and Badger, C. and Bae, S. and Bagnasco, S. and Bagui, E. and Baier, J. G. and Baiotti, L. and Bajpai, R. and Baka, T. and Ball, M. and Ballardin, G. and Ballmer, S. W. and Banagiri, S. and Banerjee, B. and Bankar, D. and Baral, P. and Barayoga, J. C. and Barish, B. C. and Barker, D. and Barneo, P. and Barone, F. and Barr, B. and Barsotti, L. and Barsuglia, M. and Barta, D. and Bartoletti, A. M. and Barton, M. A. and Bartos, I. and Basak, S. and Basalaev, A. and Bassiri, R. and Basti, A. and Bates, D. E. and Bawaj, M. and Baxi, P. and Bayley, J. C. and Baylor, A. C. and II, P. A. Baynard and Bazzan, M. and Bedakihale, V. M. and Beirnaert, F. and Bejger, M. and Belardinelli, D. and Bell, A. S. and Benedetto, V. and Benoit, W. and Bentara, I. and Bentley, J. D. and Yaala, M. Ben and Bera, S. and Berbel, M. and Bergamin, F. and Berger, B. K. and Bernuzzi, S. and Beroiz, M. and Berry, C. P. L. and Bersanetti, D. and Bertolini, A. and Betzwieser, J. and Beveridge, D. and Bevins, N. and Bhandare, R. and Bhardwaj, U. and Bhatt, R. and Bhattacharjee, D. and Bhaumik, S. and Bhowmick, S. and Bianchi, A. and Bilenko, I. A. and Billingsley, G. and Binetti, A. and Bini, S. and Birnholtz, O. and Biscoveanu, S. and Bisht, A. and Bitossi, M. and Bizouard, M.-A. and Blackburn, J. K. and Blagg, L. A. and Blair, C. D. and Blair, D. G. and Bobba, F. and Bode, N. and Boileau, G. and Boldrini, M. and Bolingbroke, G. N. and Bolliand, A. and Bonavena, L. D. and Bondarescu, R. and Bondu, F. and Bonilla, E. and Bonilla, M. S. and Bonino, A. and Bonnand, R. and Booker, P. and Borchers, A. and Boschi, V. and Bose, S. and Bossilkov, V. and Boudart, V. and Boudon, A. and Bozzi, A. and Bradaschia, C. and Brady, P. R. and Braglia, M. and Branch, A. and Branchesi, M. and Brandt, J. and Braun, I. and Breschi, M. and Briant, T. and Brillet, A. and Brinkmann, M. and Brockill, P. and Brockmueller, E. and Brooks, A. F. and Brown, B. C. and Brown, D. D. and Brozzetti, M. L. and Brunett, S. and Bruno, G. and Bruntz, R. and Bryant, J. and Bucci, F. and Buchanan, J. and Bulashenko, O. and Bulik, T. and Bulten, H. J. and Buonanno, A. and Burtnyk, K. and Buscicchio, R. and Buskulic, D. and Buy, C. and Byer, R. L. and Davies, G. S. Cabourn and Cabras, G. and Cabrita, R. and Cáceres-Barbosa, V. and Cadonati, L. and Cagnoli, G. and Cahillane, C. and Bustillo, J. Calderón and Callister, T. A. and Calloni, E. and Camp, J. B. and Canepa, M. and Santoro, G. Caneva and Cannon, K. C. and Cao, H. and Capistran, L. A. and Capocasa, E. and Capote, E. and Carapella, G. and Carbognani, F. and Carlassara, M. and Carlin, J. B. and Carpinelli, M. and Carrillo, G. and Carter, J. J. and Carullo, G. and Diaz, J. Casanueva and Casentini, C. and Castro-Lucas, S. Y. and Caudill, S. and Cavaglià, M. and Cavalieri, R. and Cella, G. and Cerdá-Durán, P. and Cesarini, E. and Chaibi, W. and Chakraborty, P. and Subrahmanya, S. Chalathadka and Chan, J. C. L. and Chan, M. and Chandra, K. and Chang, R.-J. and Chao, S. and Char, P. and Charlton, E. L. and Charlton, P. and Chassande-Mottin, E. and Chatterjee, C. and Chatterjee, Debarati and Chatterjee, Deep and Chattopadhyay, D. and Chaturvedi, M. and Chaty, S. and Chatziioannou, K. and Chen, A. and Chen, A. H.-Y. and Chen, D. and Chen, H. and Chen, H. Y. and Chen, J. and Chen, K. H. and Chen, Y. and Chen, Yanbei and Chen, Yitian and Cheng, H. P. and Chessa, P. and Cheung, H. T. and Cheung, S. Y. and Chiadini, F. and Chiarini, G. and Chierici, R. and Chincarini, A. and Chiofalo, M. L. and Chiummo, A. and Chou, C. and Choudhary, S. and Christensen, N. and Chua, S. S. Y. and Chugh, P. and Ciani, G. and Ciecielag, P. and Cieślar, M. and Cifaldi, M. and Ciolfi, R. and Clara, F. and Clark, J. A. and Clarke, J. and Clarke, T. A. and Clearwater, P. and Clesse, S. and Coccia, E. and Codazzo, E. and Cohadon, P.-F. and Colace, S. and Colleoni, M. and Collette, C. G. and Collins, J. and Colloms, S. and Colombo, A. and Colpi, M. and Compton, C. M. and Connolly, G. and Conti, L. and Corbitt, T. R. and Cordero-Carrión, I. and Corezzi, S. and Cornish, N. J. and Corsi, A. and Cortese, S. and Costa, C. A. and Cottingham, R. and Coughlin, M. W. and Couineaux, A. and Coulon, J.-P. and Countryman, S. T. and Coupechoux, J.-F. and Couvares, P. and Coward, D. M. and Cowart, M. J. and Coyne, R. and Craig, K. and Creed, R. and Creighton, J. D. E. and Creighton, T. D. and Cremonese, P. and Criswell, A. W. and Crockett-Gray, J. C. G. and Crook, S. and Crouch, R. and Csizmazia, J. and Cudell, J. R. and Cullen, T. J. and Cumming, A. and Cuoco, E. and Cusinato, M. and Dabadie, P. and Dal Canton, T. and Dall’Osso, S. and Dal Pra, S. and Dálya, G. and D’Angelo, B. and Danilishin, S. and D’Antonio, S. and Danzmann, K. and Darroch, K. E. and Dartez, L. P. and Dasgupta, A. and Datta, S. and Dattilo, V. and Daumas, A. and Davari, N. and Dave, I. and Davenport, A. and Davier, M. and Davies, T. F. and Davis, D. and Davis, L. and Davis, M. C. and Davis, P. J. and Dax, M. and De Bolle, J. and Deenadayalan, M. and Degallaix, J. and De Laurentis, M. and Deléglise, S. and De Lillo, F. and Dell’Aquila, D. and Del Pozzo, W. and De Marco, F. and De Matteis, F. and D’Emilio, V. and Demos, N. and Dent, T. and Depasse, A. and DePergola, N. and De Pietri, R. and De Rosa, R. and De Rossi, C. and DeSalvo, R. and De Simone, R. and Dhani, A. and Diab, R. and Díaz, M. C. and Di Cesare, M. and Dideron, G. and Didio, N. A. and Dietrich, T. and Di Fiore, L. and Di Fronzo, C. and Di Giovanni, M. and Di Girolamo, T. and Diksha, D. and Di Michele, A. and Ding, J. and Di Pace, S. and Di Palma, I. and Di Renzo, F. and {Divyajyoti} and Dmitriev, A. and Doctor, Z. and Dohmen, E. and Doleva, P. P. and Dominguez, D. and D’Onofrio, L. and Donovan, F. and Dooley, K. L. and Dooney, T. and Doravari, S. and Dorosh, O. and Drago, M. and Driggers, J. C. and Ducoin, J.-G. and Dunn, L. and Dupletsa, U. and D’Urso, D. and Duval, H. and Duverne, P.-A. and Dwyer, S. E. and Eassa, C. and Ebersold, M. and Eckhardt, T. and Eddolls, G. and Edelman, B. and Edo, T. B. and Edy, O. and Effler, A. and Eichholz, J. and Einsle, H. and Eisenmann, M. and Eisenstein, R. A. and Ejlli, A. and Eleveld, R. M. and Emma, M. and Endo, K. and Engl, A. J. and Enloe, E. and Errico, L. and Essick, R. C. and Estellés, H. and Estevez, D. and Etzel, T. and Evans, M. and Evstafyeva, T. and Ewing, B. E. and Ezquiaga, J. M. and Fabrizi, F. and Faedi, F. and Fafone, V. and Fairhurst, S. and Farah, A. M. and Farr, B. and Farr, W. M. and Favaro, G. and Favata, M. and Fays, M. and Fazio, M. and Feicht, J. and Fejer, M. M. and Felicetti, R. . and Fenyvesi, E. and Ferguson, D. L. and Ferraiuolo, S. and Ferrante, I. and Ferreira, T. A. and Fidecaro, F. and Figura, P. and Fiori, A. and Fiori, I. and Fishbach, M. and Fisher, R. P. and Fittipaldi, R. and Fiumara, V. and Flaminio, R. and Fleischer, S. M. and Fleming, L. S. and Floden, E. and Foley, E. M. and Fong, H. and Font, J. A. and Fornal, B. and Forsyth, P. W. F. and Franceschetti, K. and Franchini, N. and Frasca, S. and Frasconi, F. and Mascioli, A. Frattale and Frei, Z. and Freise, A. and Freitas, O. and Frey, R. and Frischhertz, W. and Fritschel, P. and Frolov, V. V. and Fronzé, G. G. and Fuentes-Garcia, M. and Fujii, S. and Fujimori, T. and Fulda, P. and Fyffe, M. and Gadre, B. and Gair, J. R. and Galaudage, S. and Galdi, V. and Gallagher, H. and Gallardo, S. and Gallego, B. and Gamba, R. and Gamboa, A. and Ganapathy, D. and Ganguly, A. and Garaventa, B. and García-Bellido, J. and García Núñez, C. and García-Quirós, C. and Gardner, J. W. and Gardner, K. A. and Gargiulo, J. and Garron, A. and Garufi, F. and Gasbarra, C. and Gateley, B. and Gayathri, V. and Gemme, G. and Gennai, A. and Gennari, V. and George, J. and George, R. and Gerberding, O. and Gergely, L. and Ghonge, S. and Ghosh, Archisman and Ghosh, Sayantan and Ghosh, Shaon and Ghosh, Shrobana and Ghosh, Suprovo and Ghosh, Tathagata and Giacoppo, L. and Giaime, J. A. and Giardina, K. D. and Gibson, D. R. and Gibson, D. T. and Gier, C. and Giri, P. and Gissi, F. and Gkaitatzis, S. and Glanzer, J. and Glotin, F. and Godfrey, J. and Godwin, P. and Goebbels, N. L. and Goetz, E. and Golomb, J. and Lopez, S. Gomez and Goncharov, B. and Gong, Y. and González, G. and Goodarzi, P. and Goode, S. and Goodwin-Jones, A. W. and Gosselin, M. and Göttel, A. S. and Gouaty, R. and Gould, D. W. and Govorkova, K. and Goyal, S. and Grace, B. and Grado, A. and Graham, V. and Granados, A. E. and Granata, M. and Granata, V. and Gras, S. and Grassia, P. and Gray, A. and Gray, C. and Gray, R. and Greco, G. and Green, A. C. and Green, S. M. and Green, S. R. and Gretarsson, A. M. and Gretarsson, E. M. and Griffith, D. and Griffiths, W. L. and Griggs, H. L. and Grignani, G. and Grimaldi, A. and Grimaud, C. and Grote, H. and Guerra, D. and Guetta, D. and Guidi, G. M. and Guimaraes, A. R. and Gulati, H. K. and Gulminelli, F. and Gunny, A. M. and Guo, H. and Guo, W. and Guo, Y. and Gupta, Anchal and Gupta, Anuradha and Gupta, Ish and Gupta, N. C. and Gupta, P. and Gupta, S. K. and Gupta, T. and Gupte, N. and Gurs, J. and Gutierrez, N. and Guzman, F. and H, H.-Y. and Haba, D. and Haberland, M. and Haino, S. and Hall, E. D. and Hamilton, E. Z. and Hammond, G. and Han, W.-B. and Haney, M. and Hanks, J. and Hanna, C. and Hannam, M. D. and Hannuksela, O. A. and Hanselman, A. G. and Hansen, H. and Hanson, J. and Harada, R. and Hardison, A. R. and Haris, K. and Harmark, T. and Harms, J. and Harry, G. M. and Harry, I. W. and Hart, J. and Haskell, B. and Haster, C.-J. and Hathaway, J. S. and Haughian, K. and Hayakawa, H. and Hayama, K. and Hayes, R. and Heffernan, A. and Heidmann, A. and Heintze, M. C. and Heinze, J. and Heinzel, J. and Heitmann, H. and Hellman, F. and Hello, P. and Helmling-Cornell, A. F. and Hemming, G. and Henderson-Sapir, O. and Hendry, M. and Heng, I. S. and Hennes, E. and Henshaw, C. and Hertog, T. and Heurs, M. and Hewitt, A. L. and Heyns, J. and Higginbotham, S. and Hild, S. and Hill, S. and Himemoto, Y. and Hirata, N. and Hirose, C. and Hoang, S. and Hochheim, S. and Hofman, D. and Holland, N. A. and Holley-Bockelmann, K. and Holmes, Z. J. and Holz, D. E. and Honet, L. and Hong, C. and Hornung, J. and Hoshino, S. and Hough, J. and Hourihane, S. and Howell, E. J. and Hoy, C. G. and Hrishikesh, C. A. and Hsieh, H.-F. and Hsiung, C. and Hsu, H. C. and Hsu, W.-F. and Hu, P. and Hu, Q. and Huang, H. Y. and Huang, Y.-J. and Huddart, A. D. and Hughey, B. and Hui, D. C. Y. and Hui, V. and Husa, S. and Huxford, R. and Huynh-Dinh, T. and Iampieri, L. and Iandolo, G. A. and Ianni, M. and Iess, A. and Imafuku, H. and Inayoshi, K. and Inoue, Y. and Iorio, G. and Iqbal, M. H. and Irwin, J. and Ishikawa, R. and Isi, M. and Ismail, M. A. and Itoh, Y. and Iwanaga, H. and Iwaya, M. and Iyer, B. R. and JaberianHamedan, V. and Jacquet, C. and Jacquet, P.-E. and Jadhav, S. J. and Jadhav, S. P. and Jain, T. and James, A. L. and James, P. A. and Jamshidi, R. and Janquart, J. and Janssens, K. and Janthalur, N. N. and Jaraba, S. and Jaranowski, P. and Jaume, R. and Javed, W. and Jennings, A. and Jia, W. and Jiang, J. and Kubisz, J. and Johanson, C. and Johns, G. R. and Johnson, N. A. and Johnson-McDaniel, N. K. and Johnston, M. C. and Johnston, R. and Johny, N. and Jones, D. H. and Jones, D. I. and Jones, R. and Jose, S. and Joshi, P. and Ju, L. and Jung, K. and Junker, J. and Juste, V. and Kajita, T. and Kaku, I. and Kalaghatgi, C. and Kalogera, V. and Kamiizumi, M. and Kanda, N. and Kandhasamy, S. and Kang, G. and Kanner, J. B. and Kapadia, S. J. and Kapasi, D. P. and Karat, S. and Karathanasis, C. and Kashyap, R. and Kasprzack, M. and Kastaun, W. and Kato, T. and Katsavounidis, E. and Katzman, W. and Kaushik, R. and Kawabe, K. and Kawamoto, R. and Kazemi, A. and Kedia, A. and Keitel, D. and Kelley-Derzon, J. and Kennington, J. and Kesharwani, R. and Key, J. S. and Khadela, R. and Khadka, S. and Khalili, F. Y. and Khan, F. and Khan, I. and Khanam, T. and Khursheed, M. and Khusid, N. M. and Kiendrebeogo, W. and Kijbunchoo, N. and Kim, C. and Kim, J. C. and Kim, K. and Kim, M. H. and Kim, S. and Kim, Y.-M. and Kimball, C. and Kinley-Hanlon, M. and Kinnear, M. and Kissel, J. S. and Klimenko, S. and Knee, A. M. and Knust, N. and Kobayashi, K. and Koch, P. and Koehlenbeck, S. M. and Koekoek, G. and Kohri, K. and Kokeyama, K. and Koley, S. and Kolitsidou, P. and Kolstein, M. and Komori, K. and Kong, A. K. H. and Kontos, A. and Korobko, M. and Kossak, R. V. and Kou, X. and Koushik, A. and Kouvatsos, N. and Kovalam, M. and Kozak, D. B. and Kranzhoff, S. L. and Kringel, V. and Krishnendu, N. V. and Królak, A. and Kruska, K. and Kuehn, G. and Kuijer, P. and Kulkarni, S. and Ramamohan, A. Kulur and Kumar, A. and Kumar, Praveen and Kumar, Prayush and Kumar, Rahul and Kumar, Rakesh and Kume, J. and Kuns, K. and Kuntimaddi, N. and Kuroyanagi, S. and Kurth, N. J. and Kuwahara, S. and Kwak, K. and Kwan, K. and Kwok, J. and Lacaille, G. and Lagabbe, P. and Laghi, D. and Lai, S. and Laity, A. H. and Lakkis, M. H. and Lalande, E. and Lalleman, M. and Lalremruati, P. C. and Landry, M. and Landry, P. and Lane, B. B. and Lang, R. N. and Lange, J. and Lantz, B. and La Rana, A. and La Rosa, I. and Lartaux-Vollard, A. and Lasky, P. D. and Lawrence, J. and Lawrence, M. N. and Laxen, M. and Lazzarini, A. and Lazzaro, C. and Leaci, P. and Lecoeuche, Y. K. and Lee, H. M. and Lee, H. W. and Lee, K. and Lee, R.-K. and Lee, R. and Lee, S. and Lee, Y. and Legred, I. N. and Lehmann, J. and Lehner, L. and Le Jean, M. and Lemaître, A. and Lenti, M. and Leonardi, M. and Lequime, M. and Leroy, N. and Lesovsky, M. and Letendre, N. and Lethuillier, M. and Levin, S. E. and Levin, Y. and Leyde, K. and Li, A. K. Y. and Li, K. L. and Li, T. G. F. and Li, X. and Li, Z. and Lihos, A. and Lin, C-Y. and Lin, C.-Y. and Lin, E. T. and Lin, F. and Lin, H. and Lin, L. C.-C. and Lin, Y.-C. and Linde, F. and Linker, S. D. and Littenberg, T. B. and Liu, A. and Liu, G. C. and Liu, Jian and Villarreal, F. Llamas and Llobera-Querol, J. and Lo, R. K. L. and Locquet, J.-P. and London, L. T. and Longo, A. and Lopez, D. and Lopez Portilla, M. and Lorenzini, M. and Lorenzo-Medina, A. and Loriette, V. and Lormand, M. and Losurdo, G. and IV, T. P. Lott and Lough, J. D. and Loughlin, H. A. and Lousto, C. O. and Lowry, M. J. and Lu, N. and Lück, H. and Lumaca, D. and Lundgren, A. P. and Lussier, A. W. and Ma, L.-T. and Ma, S. and Ma’arif, M. and Macas, R. and Macedo, A. and MacInnis, M. and Maciy, R. R. and Macleod, D. M. and MacMillan, I. A. O. and Macquet, A. and Macri, D. and Maeda, K. and Maenaut, S. and Hernandez, I. Magaña and Magare, S. S. and Magazzù, C. and Magee, R. M. and Maggio, E. and Maggiore, R. and Magnozzi, M. and Mahesh, M. and Mahesh, S. and Maini, M. and Majhi, S. and Majorana, E. and Makarem, C. N. and Makelele, E. and Malaquias-Reis, J. A. and Mali, U. and Maliakal, S. and Malik, A. and Man, N. and Mandic, V. and Mangano, V. and Mannix, B. and Mansell, G. L. and Mansingh, G. and Manske, M. and Mantovani, M. and Mapelli, M. and Marchesoni, F. and Marín Pina, D. and Marion, F. and Márka, S. and Márka, Z. and Markosyan, A. S. and Markowitz, A. and Maros, E. and Marsat, S. and Martelli, F. and Martin, I. W. and Martin, R. M. and Martinez, B. B. and Martinez, M. and Martinez, V. and Martini, A. and Martinovic, K. and Martins, J. C. and Martynov, D. V. and Marx, E. J. and Massaro, L. and Masserot, A. and Masso-Reid, M. and Mastrodicasa, M. and Mastrogiovanni, S. and Matcovich, T. and Matiushechkina, M. and Matsuyama, M. and Mavalvala, N. and Maxwell, N. and McCarrol, G. and McCarthy, R. and McClelland, D. E. and McCormick, S. and McCuller, L. and McEachin, S. and McElhenny, C. and McGhee, G. I. and McGinn, J. and McGowan, K. B. M. and McIver, J. and McLeod, A. and McRae, T. and Meacher, D. and Meijer, Q. and Melatos, A. and Mellaerts, S. and Menendez-Vazquez, A. and Menoni, C. S. and Mera, F. and Mercer, R. A. and Mereni, L. and Merfeld, K. and Merilh, E. L. and Mérou, J. R. and Merritt, J. D. and Merzougui, M. and Messenger, C. and Messick, C. and Meyer-Conde, M. and Meylahn, F. and Mhaske, A. and Miani, A. and Miao, H. and Michaloliakos, I. and Michel, C. and Michimura, Y. and Middleton, H. and Miller, A. L. and Miller, S. and Millhouse, M. and Milotti, E. and Milotti, V. and Minenkov, Y. and Mio, N. and Mir, Ll. M. and Mirasola, L. and Miravet-Tenés, M. and Miritescu, C.-A. and Mishra, A. K. and Mishra, A. and Mishra, C. and Mishra, T. and Mitchell, A. L. and Mitchell, J. G. and Mitra, S. and Mitrofanov, V. P. and Mittleman, R. and Miyakawa, O. and Miyamoto, S. and Miyoki, S. and Mo, G. and Mobilia, L. and Mohapatra, S. R. P. and Mohite, S. R. and Molina-Ruiz, M. and Mondal, C. and Mondin, M. and Montani, M. and Moore, C. J. and Moraru, D. and More, A. and More, S. and Moreno, G. and Morgan, C. and Morisaki, S. and Moriwaki, Y. and Morras, G. and Moscatello, A. and Mourier, P. and Mours, B. and Mow-Lowry, C. M. and Muciaccia, F. and Mukherjee, Arunava and Mukherjee, D. and Mukherjee, Samanwaya and Mukherjee, Soma and Mukherjee, Subroto and Mukherjee, Suvodip and Mukund, N. and Mullavey, A. and Munch, J. and Mundi, J. and Mungioli, C. L. and Oberg, W. R. Munn and Murakami, Y. and Murakoshi, M. and Murray, P. G. and Muusse, S. and Nabari, D. and Nadji, S. L. and Nagar, A. and Nagarajan, N. and Nagler, K. N. and Nakagaki, K. and Nakamura, K. and Nakano, H. and Nakano, M. and Nandi, D. and Napolano, V. and Narayan, P. and Nardecchia, I. and Narikawa, T. and Narola, H. and Naticchioni, L. and Nayak, R. K. and Neilson, J. and Nelson, A. and Nelson, T. J. N. and Nery, M. and Neunzert, A. and Ng, S. and Quynh, L. Nguyen and Nichols, S. A. and Nielsen, A. B. and Nieradka, G. and Niko, A. and Nishino, Y. and Nishizawa, A. and Nissanke, S. and Nitoglia, E. and Niu, W. and Nocera, F. and Norman, M. and North, C. and Novak, J. and Siles, J. F. Nuño and Nuttall, L. K. and Obayashi, K. and Oberling, J. and O’Dell, J. and Oertel, M. and Offermans, A. and Oganesyan, G. and Oh, J. J. and Oh, K. and O’Hanlon, T. and Ohashi, M. and Ohkawa, M. and Ohme, F. and Oliveira, A. S. and Oliveri, R. and O’Neal, B. and Oohara, K. and O’Reilly, B. and Ormsby, N. D. and Orselli, M. and O’Shaughnessy, R. and O’Shea, S. and Oshima, Y. and Oshino, S. and Ossokine, S. and Osthelder, C. and Ota, I. and Ottaway, D. J. and Ouzriat, A. and Overmier, H. and Owen, B. J. and Pace, A. E. and Pagano, R. and Page, M. A. and Pai, A. and Pal, A. and Pal, S. and Palaia, M. A. and Pálfi, M. and Palma, P. P. and Palomba, C. and Palud, P. and Pan, H. and Pan, J. and Pan, K. C. and Panai, R. and Panda, P. K. and Pandey, S. and Panebianco, L. and Pang, P. T. H. and Pannarale, F. and Pannone, K. A. and Pant, B. C. and Panther, F. H. and Paoletti, F. and Paolone, A. and Papalexakis, E. E. and Papalini, L. and Papigkiotis, G. and Paquis, A. and Parisi, A. and Park, B.-J. and Park, J. and Parker, W. and Pascale, G. and Pascucci, D. and Pasqualetti, A. and Passaquieti, R. and Passenger, L. and Passuello, D. and Patane, O. and Pathak, D. and Pathak, M. and Patra, A. and Patricelli, B. and Patron, A. S. and Paul, K. and Paul, S. and Payne, E. and Pearce, T. and Pedraza, M. and Pegna, R. and Pele, A. and Arellano, F. E. Peña and Penn, S. and Penuliar, M. D. and Perego, A. and Pereira, Z. and Perez, J. J. and Périgois, C. and Perna, G. and Perreca, A. and Perret, J. and Perriès, S. and Perry, J. W. and Pesios, D. and Petracca, S. and Petrillo, C. and Pfeiffer, H. P. and Pham, H. and Pham, K. A. and Phukon, K. S. and Phurailatpam, H. and Piarulli, M. and Piccari, L. and Piccinni, O. J. and Pichot, M. and Piendibene, M. and Piergiovanni, F. and Pierini, L. and Pierra, G. and Pierro, V. and Pietrzak, M. and Pillas, M. and Pilo, F. and Pinard, L. and Pinto, I. M. and Pinto, M. and Piotrzkowski, B. J. and Pirello, M. and Pitkin, M. D. and Placidi, A. and Placidi, E. and Planas, M. L. and Plastino, W. and Poggiani, R. and Polini, E. and Pompili, L. and Poon, J. and Porcelli, E. and Porter, E. K. and Posnansky, C. and Poulton, R. and Powell, J. and Pracchia, M. and Pradhan, B. K. and Pradier, T. and Prajapati, A. K. and Prasai, K. and Prasanna, R. and Prasia, P. and Pratten, G. and Principe, G. and Principe, M. and Prodi, G. A. and Prokhorov, L. and Prosposito, P. and Puecher, A. and Pullin, J. and Punturo, M. and Puppo, P. and Pürrer, M. and Qi, H. and Qin, J. and Quéméner, G. and Quetschke, V. and Quigley, C. and Quinonez, P. J. and Raab, F. J. and Raabith, S. S. and Raaijmakers, G. and Raja, S. and Rajan, C. and Rajbhandari, B. and Ramirez, K. E. and Vidal, F. A. Ramis and Ramos-Buades, A. and Rana, D. and Ranjan, S. and Ransom, K. and Rapagnani, P. and Ratto, B. and Rawat, S. and Ray, A. and Raymond, V. and Razzano, M. and Read, J. and Payo, M. Recaman and Regimbau, T. and Rei, L. and Reid, S. and Reitze, D. H. and Relton, P. and Renzini, A. I. and Rettegno, P. and Revenu, B. and Reyes, R. and Rezaei, A. S. and Ricci, F. and Ricci, M. and Ricciardone, A. and Richardson, J. W. and Richardson, M. and Rijal, A. and Riles, K. and Riley, H. K. and Rinaldi, S. and Rittmeyer, J. and Robertson, C. and Robinet, F. and Robinson, M. and Rocchi, A. and Rolland, L. and Rollins, J. G. and Romano, A. E. and Romano, R. and Romero, A. and Romero-Shaw, I. M. and Romie, J. H. and Ronchini, S. and Roocke, T. J. and Rosa, L. and Rosauer, T. J. and Rose, C. A. and Rosińska, D. and Ross, M. P. and Rossello, M. and Rowan, S. and Roy, S. K. and Roy, S. and Rozza, D. and Ruggi, P. and Ruhama, N. and Morales, E. Ruiz and Ruiz-Rocha, K. and Sachdev, S. and Sadecki, T. and Sadiq, J. and Saffarieh, P. and Sah, M. R. and Saha, S. S. and Saha, S. and Sainrat, T. and Menon, S. Sajith and Sakai, K. and Sakellariadou, M. and Sakon, S. and Salafia, O. S. and Salces-Carcoba, F. and Salconi, L. and Saleem, M. and Salemi, F. and Sallé, M. and Salvador, S. and Sanchez, A. and Sanchez, E. J. and Sanchez, J. H. and Sanchez, L. E. and Sanchis-Gual, N. and Sanders, J. R. and Sänger, E. M. and Santoliquido, F. and Saravanan, T. R. and Sarin, N. and Sasaoka, S. and Sasli, A. and Sassi, P. and Sassolas, B. and Satari, H. and Sathyaprakash, B. S. and Sato, R. and Sato, Y. and Sauter, O. and Savage, R. L. and Sawada, T. and Sawant, H. L. and Sayah, S. and Scacco, V. and Schaetzl, D. and Scheel, M. and Schiebelbein, A. and Schiworski, M. G. and Schmidt, P. and Schmidt, S. and Schnabel, R. and Schneewind, M. and Schofield, R. M. S. and Schouteden, K. and Schulte, B. W. and Schutz, B. F. and Schwartz, E. and Scialpi, M. and Scott, J. and Scott, S. M. and Seetharamu, T. C. and Seglar-Arroyo, M. and Sekiguchi, Y. and Sellers, D. and Sengupta, A. S. and Sentenac, D. and Seo, E. G. and Seo, J. W. and Sequino, V. and Serra, M. and Servignat, G. and Sevrin, A. and Shaffer, T. and Shah, U. S. and Shaikh, M. A. and Shao, L. and Sharma, A. K. and Sharma, P. and Sharma-Chaudhary, S. and Shaw, M. R. and Shawhan, P. and Shcheblanov, N. S. and Sheridan, E. and Shikano, Y. and Shikauchi, M. and Shimode, K. and Shinkai, H. and Shiota, J. and Shoemaker, D. H. and Shoemaker, D. M. and Short, R. W. and ShyamSundar, S. and Sider, A. and Siegel, H. and Sieniawska, M. and Sigg, D. and Silenzi, L. and Simmonds, M. and Singer, L. P. and Singh, A. and Singh, D. and Singh, M. K. and Singh, S. and Singha, A. and Sintes, A. M. and Sipala, V. and Skliris, V. and Slagmolen, B. J. J. and Slaven-Blair, T. J. and Smetana, J. and Smith, J. R. and Smith, L. and Smith, R. J. E. and Smith, W. J. and Soldateschi, J. and Somiya, K. and Song, I. and Soni, K. and Soni, S. and Sordini, V. and Sorrentino, F. and Sorrentino, N. and Sotani, H. and Soulard, R. and Southgate, A. and Spagnuolo, V. and Spencer, A. P. and Spera, M. and Spinicelli, P. and Spoon, J. B. and Sprague, C. A. and Srivastava, A. K. and Stachurski, F. and Steer, D. A. and Steinlechner, J. and Steinlechner, S. and Stergioulas, N. and Stevens, P. and Stevenson, S. and StPierre, M. and Stratta, G. and Strong, M. D. and Strunk, A. and Sturani, R. and Stuver, A. L. and Suchenek, M. and Sudhagar, S. and Sueltmann, N. and Suleiman, L. and Sullivan, K. D. and Sun, L. and Sunil, S. and Suresh, J. and Sutton, P. J. and Suzuki, T. and Suzuki, Y. and Swinkels, B. L. and Syx, A. and Szczepańczyk, M. J. and Szewczyk, P. and Tacca, M. and Tagoshi, H. and Tait, S. C. and Takahashi, H. and Takahashi, R. and Takamori, A. and Takase, T. and Takatani, K. and Takeda, H. and Takeshita, K. and Talbot, C. and Tamaki, M. and Tamanini, N. and Tanabe, D. and Tanaka, K. and Tanaka, S. J. and Tanaka, T. and Tang, D. and Tanioka, S. and Tanner, D. B. and Tao, L. and Tapia, R. D. and San Martín, E. N. Tapia and Tarafder, R. and Taranto, C. and Taruya, A. and Tasson, J. D. and Teloi, M. and Tenorio, R. and Themann, H. and Theodoropoulos, A. and Thirugnanasambandam, M. P. and Thomas, L. M. and Thomas, M. and Thomas, P. and Thompson, J. E. and Thondapu, S. R. and Thorne, K. A. and Thrane, E. and Tissino, J. and Tiwari, A. and Tiwari, P. and Tiwari, S. and Tiwari, V. and Todd, M. R. and Toivonen, A. M. and Toland, K. and Tolley, A. E. and Tomaru, T. and Tomita, K. and Tomura, T. and Tong, H. and Tong-Yu, C. and Toriyama, A. and Toropov, N. and Torres-Forné, A. and Torrie, C. I. and Toscani, M. and e Melo, I. Tosta and Tournefier, E. and Trapananti, A. and Travasso, F. and Traylor, G. and Trevor, M. and Tringali, M. C. and Tripathee, A. and Troian, G. and Troiano, L. and Trovato, A. and Trozzo, L. and Trudeau, R. J. and Tsang, T. T. L. and Tso, R. and Tsuchida, S. and Tsukada, L. and Tsutsui, T. and Turbang, K. and Turconi, M. and Turski, C. and Ubach, H. and Uchikata, N. and Uchiyama, T. and Udall, R. P. and Uehara, T. and Uematsu, M. and Ueno, K. and Ueno, S. and Undheim, V. and Ushiba, T. and Vacatello, M. and Vahlbruch, H. and Vaidya, N. and Vajente, G. and Vajpeyi, A. and Valdes, G. and Valencia, J. and Valentini, M. and Vallejo-Peña, S. A. and Vallero, S. and Valsan, V. and van Bakel, N. and van Beuzekom, M. and van Dael, M. and van den Brand, J. F. J. and Broeck, C. Van Den and Vander-Hyde, D. C. and van der Sluys, M. and Van de Walle, A. and van Dongen, J. and Vandra, K. and van Haevermaet, H. and van Heijningen, J. V. and Van Hove, P. and VanKeuren, M. and Vanosky, J. and van Putten, M. H. P. M. and van Ranst, Z. and van Remortel, N. and Vardaro, M. and Vargas, A. F. and Varghese, J. J. and Varma, V. and Vasúth, M. and Vecchio, A. and Vedovato, G. and Veitch, J. and Veitch, P. J. and Venikoudis, S. and Venneberg, J. and Verdier, P. and Verkindt, D. and Verma, B. and Verma, P. and Verma, Y. and Vermeulen, S. M. and Vetrano, F. and Veutro, A. and Vibhute, A. M. and Viceré, A. and Vidyant, S. and Viets, A. D. and Vijaykumar, A. and Vilkha, A. and Villa-Ortega, V. and Vincent, E. T. and Vinet, J.-Y. and Viret, S. and Virtuoso, A. and Vitale, S. and Vives, A. and Vocca, H. and Voigt, D. and von Reis, E. R. G. and von Wrangel, J. S. A. and Vyatchanin, S. P. and Wade, L. E. and Wade, M. and Wagner, K. J. and Wajid, A. and Walker, M. and Wallace, G. S. and Wallace, L. and Wang, H. and Wang, J. Z. and Wang, W. H. and Wang, Z. and Waratkar, G. and Warner, J. and Was, M. and Washimi, T. and Washington, N. Y. and Watarai, D. and Wayt, K. E. and Weaver, B. R. and Weaver, B. and Weaving, C. R. and Webster, S. A. and Weinert, M. and Weinstein, A. J. and Weiss, R. and Wellmann, F. and Wen, L. and Weßels, P. and Wette, K. and Whelan, J. T. and Whiting, B. F. and Whittle, C. and Wildberger, J. B. and Wilk, O. S. and Wilken, D. and Wilkin, A. T. and Willadsen, D. J. and Willetts, K. and Williams, D. and Williams, M. J. and Williams, N. S. and Willis, J. L. and Willke, B. and Wils, M. and Winterflood, J. and Wipf, C. C. and Woan, G. and Woehler, J. and Wofford, J. K. and Wolfe, N. E. and Wong, H. T. and Wong, H. W. Y. and Wong, I. C. F. and Wright, J. L. and Wright, M. and Wu, C. and Wu, D. S. and Wu, H. and Wuchner, E. and Wysocki, D. M. and Xu, V. A. and Xu, Y. and Yadav, N. and Yamamoto, H. and Yamamoto, K. and Yamamoto, T. S. and Yamamoto, T. and Yamamura, S. and Yamazaki, R. and Yan, S. and Yan, T. and Yang, F. W. and Yang, F. and Yang, K. Z. and Yang, Y. and Yarbrough, Z. and Yasui, H. and Yeh, S.-W. and Yelikar, A. B. and Yin, X. and Yokoyama, J. and Yokozawa, T. and Yoo, J. and Yu, H. and Yuan, S. and Yuzurihara, H. and Zadrożny, A. and Zanolin, M. and Zeeshan, M. and Zelenova, T. and Zendri, J.-P. and Zeoli, M. and Zerrad, M. and Zevin, M. and Zhang, A. C. and Zhang, L. and Zhang, R. and Zhang, T. and Zhang, Y. and Zhao, C. and Zhao, Yue and Zhao, Yuhang and Zheng, Y. and Zhong, H. and Zhou, R. and Zhu, X.-J. and Zhu, Z.-H. and Zimmerman, A. B. and Zucker, M. E. and Zweizig, J. and The LIGO Scientific Collaboration, the Virgo Collaboration,  and Collaboration, the KAGRA},
	month = jul,
	year = {2024},
	pages = {L34},
}

@article{ahumada2022Search,
	title = {In {Search} of {Short} {Gamma}-{Ray} {Burst} {Optical} {Counterparts} with the {Zwicky} {Transient} {Facility}},
	volume = {932},
	issn = {0004-637X},
	url = {https://dx.doi.org/10.3847/1538-4357/ac6c29},
	doi = {10.3847/1538-4357/ac6c29},
	language = {en},
	number = {1},
	urldate = {2024-01-10},
	journal = {The Astrophysical Journal},
	publisher = {The American Astronomical Society},
	author = {Ahumada, Tomás and Anand, Shreya and Coughlin, Michael W. and Andreoni, Igor and Kool, Erik C. and Kumar, Harsh and Reusch, Simeon and Sagués-Carracedo, Ana and Stein, Robert and Cenko, S. Bradley and Kasliwal, Mansi M. and Singer, Leo P. and Dunwoody, Rachel and Mangan, Joseph and Bhalerao, Varun and Bulla, Mattia and Burns, Eric and Graham, Matthew J. and Kaplan, David L. and Perley, Daniel and Almualla, Mouza and Bloom, Joshua S. and Cunningham, Virginia and De, Kishalay and Gatkine, Pradip and Ho, Anna Y. Q. and Karambelkar, Viraj and Kong, Albert K. H. and Yao, Yuhan and Anupama, G. C. and Barway, Sudhanshu and Ghosh, Shaon and Itoh, Ryosuke and McBreen, Sheila and Bellm, Eric C. and Fremling, Christoffer and Laher, Russ R. and Mahabal, Ashish A. and Riddle, Reed L. and Rosnet, Philippe and Rusholme, Ben and Smith, Roger and Sollerman, Jesper and Bissaldi, Elisabetta and Fletcher, Corinne and Hamburg, Rachel and Mailyan, Bagrat and Malacaria, Christian and Roberts, Oliver},
	month = jun,
	year = {2022},
	pages = {40},
}

@article{rastinejad2021Probing,
	title = {Probing {Kilonova} {Ejecta} {Properties} {Using} a {Catalog} of {Short} {Gamma}-{Ray} {Burst} {Observations}},
	volume = {916},
	issn = {0004-637X},
	url = {https://doi.org/10.3847/1538-4357/ac04b4},
	doi = {10.3847/1538-4357/ac04b4},
	language = {en},
	number = {2},
	urldate = {2026-03-16},
	journal = {The Astrophysical Journal},
	publisher = {The American Astronomical Society},
	author = {Rastinejad, J. C. and Fong, W. and Kilpatrick, C. D. and Paterson, K. and Tanvir, N. R. and Levan, A. J. and Metzger, B. D. and Berger, E. and Chornock, R. and Cobb, B. E. and Laskar, T. and Milne, P. and Nugent, A. E. and Smith, N.},
	month = aug,
	year = {2021},
	pages = {89},
}

@article{ivezic2019LSST,
	title = {{LSST}: {From} {Science} {Drivers} to {Reference} {Design} and {Anticipated} {Data} {Products}},
	language = {en},
	journal = {The Astrophysical Journal},
	author = {Ivezic, Zeljko},
	year = {2019},
}

@article{cutler1994Gravitational,
	title = {Gravitational waves from merging compact binaries: {How} accurately can one extract the binary's parameters from the inspiral waveform?},
	volume = {49},
	shorttitle = {Gravitational waves from merging compact binaries},
	url = {https://link.aps.org/doi/10.1103/PhysRevD.49.2658},
	doi = {10.1103/PhysRevD.49.2658},
	number = {6},
	urldate = {2023-09-14},
	journal = {Physical Review D},
	publisher = {American Physical Society},
	author = {Cutler, Curt and Flanagan, Eanna E.},
	month = mar,
	year = {1994},
	pages = {2658--2697},
}

@article{villar2017Combined,
	title = {The {Combined} {Ultraviolet}, {Optical}, and {Near}-{Infrared} {Light} {Curves} of the {Kilonova} {Associated} with the {Binary} {Neutron} {Star} {Merger} {GW170817}: {Unified} {Data} {Set}, {Analytic} {Models}, and {Physical} {Implications}},
	volume = {851},
	issn = {2041-8205, 2041-8213},
	shorttitle = {The {Combined} {Ultraviolet}, {Optical}, and {Near}-{Infrared} {Light} {Curves} of the {Kilonova} {Associated} with the {Binary} {Neutron} {Star} {Merger} {GW170817}},
	url = {http://arxiv.org/abs/1710.11576},
	doi = {10.3847/2041-8213/aa9c84},
	number = {1},
	urldate = {2026-03-14},
	journal = {The Astrophysical Journal Letters},
	author = {Villar, V. Ashley and Guillochon, James and Berger, Edo and Metzger, Brian D. and Cowperthwaite, Philip S. and Nicholl, Matt and Alexander, Kate D. and Blanchard, Peter K. and Chornock, Ryan and Eftekhari, Tarraneh and Fong, Wen-fai and Margutti, Raffaella and Williams, Peter K. G.},
	month = dec,
	year = {2017},
	note = {arXiv:1710.11576 [astro-ph]},
	pages = {L21},
}

@article{meegan2009Fermi,
	title = {The {Fermi} {Gamma}-ray {Burst} {Monitor}},
	volume = {702},
	issn = {0004-637X},
	url = {https://ui.adsabs.harvard.edu/abs/2009ApJ...702..791M},
	doi = {10.1088/0004-637X/702/1/791},
	urldate = {2026-03-13},
	journal = {The Astrophysical Journal},
	publisher = {IOP},
	author = {Meegan, Charles and Lichti, Giselher and Bhat, P. N. and Bissaldi, Elisabetta and Briggs, Michael S. and Connaughton, Valerie and Diehl, Roland and Fishman, Gerald and Greiner, Jochen and Hoover, Andrew S. and van der Horst, Alexander J. and von Kienlin, Andreas and Kippen, R. Marc and Kouveliotou, Chryssa and McBreen, Sheila and Paciesas, W. S. and Preece, Robert and Steinle, Helmut and Wallace, Mark S. and Wilson, Robert B. and Wilson-Hodge, Colleen},
	month = sep,
	year = {2009},
	note = {ADS Bibcode: 2009ApJ...702..791M},
	pages = {791--804},
}

@article{barthelmy2005Burst,
	title = {The {Burst} {Alert} {Telescope} ({BAT}) on the {SWIFT} {Midex} {Mission}},
	volume = {120},
	issn = {1572-9672},
	url = {https://doi.org/10.1007/s11214-005-5096-3},
	doi = {10.1007/s11214-005-5096-3},
	language = {en},
	number = {3},
	urldate = {2026-03-13},
	journal = {Space Science Reviews},
	author = {Barthelmy, Scott D. and Barbier, Louis M. and Cummings, Jay R. and Fenimore, Ed E. and Gehrels, Neil and Hullinger, Derek and Krimm, Hans A. and Markwardt, Craig B. and Palmer, David M. and Parsons, Ann and Sato, Goro and Suzuki, Masaya and Takahashi, Tadayuki and Tashiro, Makota and Tueller, Jack},
	month = oct,
	year = {2005},
	pages = {143--164},
}

@article{capaccioli2011VLT,
	title = {The {VLT} {Survey} {Telescope} {Opens} to the {Sky}: {History} of a {Commissioning}},
	volume = {146},
	issn = {0722-6691},
	shorttitle = {The {VLT} {Survey} {Telescope} {Opens} to the {Sky}},
	url = {https://ui.adsabs.harvard.edu/abs/2011Msngr.146....2C},
	urldate = {2026-03-13},
	journal = {The Messenger},
	author = {Capaccioli, M. and Schipani, P.},
	month = dec,
	year = {2011},
	note = {ADS Bibcode: 2011Msngr.146....2C},
	pages = {2--7},
}

@article{broekgaarden2021Impact,
	title = {Impact of massive binary star and cosmic evolution on gravitational wave observations {I}: black hole–neutron star mergers},
	volume = {508},
	issn = {0035-8711},
	shorttitle = {Impact of massive binary star and cosmic evolution on gravitational wave observations {I}},
	url = {https://doi.org/10.1093/mnras/stab2716},
	doi = {10.1093/mnras/stab2716},
	number = {4},
	urldate = {2026-03-13},
	journal = {Monthly Notices of the Royal Astronomical Society},
	author = {Broekgaarden, Floor S and Berger, Edo and Neijssel, Coenraad J and Vigna-Gómez, Alejandro and Chattopadhyay, Debatri and Stevenson, Simon and Chruslinska, Martyna and Justham, Stephen and de Mink, Selma E and Mandel, Ilya},
	month = dec,
	year = {2021},
	pages = {5028--5063},
}

@article{lorimer2008Binary,
	title = {Binary and {Millisecond} {Pulsars}},
	volume = {11},
	issn = {2367-3613, 1433-8351},
	url = {http://arxiv.org/abs/0811.0762},
	doi = {10.12942/lrr-2008-8},
	number = {1},
	urldate = {2026-01-14},
	journal = {Living Reviews in Relativity},
	author = {Lorimer, D. R.},
	month = dec,
	year = {2008},
	note = {arXiv:0811.0762 [astro-ph]},
	pages = {8},
}

@article{abbott2017Multimessenger,
	title = {Multi-messenger {Observations} of a {Binary} {Neutron} {Star} {Merger} *},
	volume = {848},
	url = {https://doi.org/10.3847/2041-8213/aa91c9},
	doi = {10.3847/2041-8213/aa91c9},
	journal = {The Astrophysical Journal Letters},
	author = {Abbott, B. P. and Abbott, R. and Abbott, T. D. and Abernathy, M. R. and Acernese, F. and Ackley, K. and Adams, C. and Adams, T. and Addesso, P. and Adhikari, R. X. and Adya, V. B. and Affeldt, C. and Agathos, M. and Agatsuma, K. and Aggarwal, N. and Aguiar, O. D. and Aiello, L. and Ain, A. and Ajith, P. and Allen, B. and Allocca, A. and Altin, P. A. and Anderson, S. B. and Anderson, W. G. and Arai, K. and Araya, M. C. and Arceneaux, C. C. and Areeda, J. S. and Arnaud, N. and Arun, K. G. and Ascenzi, S. and Ashton, G. and Ast, M. and Aston, S. M. and Astone, P. and Aufmuth, P. and Aulbert, C. and Babak, S. and Bacon, P. and Bader, M. K.M. and Baker, P. T. and Baldaccini, F. and Ballardin, G. and Ballmer, S. W. and Barayoga, J. C. and Barclay, S. E. and Barish, B. C. and Barker, D. and Barone, F. and Barr, B. and Barsotti, L. and Barsuglia, M. and Chen, H. Y. and Chen, Y. and Cheng, C. and Chincarini, A. and Chiummo, A. and Cho, H. S. and Cho, M. and Chow, J. H. and Christensen, N. and Chu, Q. and Chua, S. and Chung, S. and Ciani, G. and Clara, F. and Clark, J. A. and Cleva, F. and Coccia, E. and Cohadon, P. F. and Colla, A. and Collette, C. G. and Cominsky, L. and Constancio, M. and Conte, A. and Conti, L. and Cook, D. and Corbitt, T. R. and Zhou, M. and Zhou, Z. and Zhu, X. J. and Zucker, M. E. and Zuraw, S. E. and Zweizig, J. and Boyle, M. and Brügmann, B. and Campanelli, M. and Clark, M. and Hamberger, D. and Kidder, L. E. and Kinsey, M. and Laguna, P. and Ossokine, S. and Scheel, M. A. and Szilagyi, B. and Teukolsky, S. and Zlochower, Y.},
	year = {2017},
	pages = {L12},
}

@article{dellavalle2006Enigmatic,
	title = {An enigmatic long-lasting γ-ray burst not accompanied by a bright supernova},
	volume = {444},
	issn = {0028-0836},
	url = {https://ui.adsabs.harvard.edu/abs/2006Natur.444.1050D},
	doi = {10.1038/nature05374},
	urldate = {2024-03-28},
	journal = {Nature},
	author = {Della Valle, M. and Chincarini, G. and Panagia, N. and Tagliaferri, G. and Malesani, D. and Testa, V. and Fugazza, D. and Campana, S. and Covino, S. and Mangano, V. and Antonelli, L. A. and D'Avanzo, P. and Hurley, K. and Mirabel, I. F. and Pellizza, L. J. and Piranomonte, S. and Stella, L.},
	month = dec,
	year = {2006},
	note = {ADS Bibcode: 2006Natur.444.1050D},
	pages = {1050--1052},
}

@article{gal-yam2006Novel,
	title = {A novel explosive process is required for the γ-ray burst {GRB} 060614},
	volume = {444},
	issn = {0028-0836},
	url = {https://ui.adsabs.harvard.edu/abs/2006Natur.444.1053G},
	doi = {10.1038/nature05373},
	urldate = {2024-03-28},
	journal = {Nature},
	author = {Gal-Yam, A. and Fox, D. B. and Price, P. A. and Ofek, E. O. and Davis, M. R. and Leonard, D. C. and Soderberg, A. M. and Schmidt, B. P. and Lewis, K. M. and Peterson, B. A. and Kulkarni, S. R. and Berger, E. and Cenko, S. B. and Sari, R. and Sharon, K. and Frail, D. and Moon, D. -S. and Brown, P. J. and Cucchiara, A. and Harrison, F. and Piran, T. and Persson, S. E. and McCarthy, P. J. and Penprase, B. E. and Chevalier, R. A. and MacFadyen, A. I.},
	month = dec,
	year = {2006},
	note = {ADS Bibcode: 2006Natur.444.1053G},
	pages = {1053--1055},
}

@article{stratta2007Study,
	title = {A study of the prompt and afterglow emission of the short {GRB} 061201},
	volume = {474},
	issn = {0004-6361},
	url = {https://ui.adsabs.harvard.edu/abs/2007A&A...474..827S},
	doi = {10.1051/0004-6361:20078006},
	urldate = {2024-03-28},
	journal = {Astronomy and Astrophysics},
	author = {Stratta, G. and D'Avanzo, P. and Piranomonte, S. and Cutini, S. and Preger, B. and Perri, M. and Conciatore, M. L. and Covino, S. and Stella, L. and Guetta, D. and Marshall, F. E. and Holland, S. T. and Stamatikos, M. and Guidorzi, C. and Mangano, V. and Antonelli, L. A. and Burrows, D. and Campana, S. and Capalbi, M. and Chincarini, G. and Cusumano, G. and D'Elia, V. and Evans, P. A. and Fiore, F. and Fugazza, D. and Giommi, P. and Osborne, J. P. and La Parola, V. and Mineo, T. and Moretti, A. and Page, K. L. and Romano, P. and Tagliaferri, G.},
	month = nov,
	year = {2007},
	note = {ADS Bibcode: 2007A\&A...474..827S},
	pages = {827--835},
}

@article{yang2015Possible,
	title = {A possible macronova in the late afterglow of the long-short burst {GRB} 060614},
	volume = {6},
	issn = {2041-1723},
	url = {https://ui.adsabs.harvard.edu/abs/2015NatCo...6.7323Y},
	doi = {10.1038/ncomms8323},
	urldate = {2024-03-28},
	journal = {Nature Communications},
	author = {Yang, Bin and Jin, Zhi-Ping and Li, Xiang and Covino, Stefano and Zheng, Xian-Zhong and Hotokezaka, Kenta and Fan, Yi-Zhong and Piran, Tsvi and Wei, Da-Ming},
	month = jun,
	year = {2015},
	note = {ADS Bibcode: 2015NatCo...6.7323Y},
	pages = {7323},
}

@article{tanvir2013Kilonova,
	title = {A "kilonova" associated with short-duration gamma-ray burst {130603B}},
	volume = {500},
	issn = {0028-0836, 1476-4687},
	url = {http://arxiv.org/abs/1306.4971},
	doi = {10.1038/nature12505},
	number = {7464},
	urldate = {2024-03-27},
	journal = {Nature},
	author = {Tanvir, N. R. and Levan, A. J. and Fruchter, A. S. and Hjorth, J. and Hounsell, R. A. and Wiersema, K. and Tunnicliffe, R.},
	month = aug,
	year = {2013},
	note = {arXiv:1306.4971 [astro-ph]},
	pages = {547--549},
}

@article{smartt2017Kilonova,
	title = {A kilonova as the electromagnetic counterpart to a gravitational-wave source},
	volume = {551},
	issn = {0028-0836, 1476-4687},
	url = {http://arxiv.org/abs/1710.05841},
	doi = {10.1038/nature24303},
	number = {7678},
	urldate = {2024-03-27},
	journal = {Nature},
	author = {Smartt, S. J. and Chen, T.-W. and Jerkstrand, A. and Coughlin, M. and Kankare, E. and Sim, S. A. and Fraser, M. and Inserra, C. and Maguire, K. and Chambers, K. C. and Huber, M. E. and Kruhler, T. and Leloudas, G. and Magee, M. and Shingles, L. J. and Smith, K. W. and Young, D. R. and Tonry, J. and Kotak, R. and Gal-Yam, A. and Lyman, J. D. and Homan, D. S. and Agliozzo, C. and Anderson, J. P. and Ashall, C. R. Angus C. and Barbarino, C. and Bauer, F. E. and Berton, M. and Botticella, M. T. and Bulla, M. and Bulger, J. and Cannizzaro, G. and Cano, Z. and Cartier, R. and Cikota, A. and Clark, P. and De Cia, A. and Della Valle, M. and Denneau, L. and Dennefeld, M. and Dessart, L. and Dimitriadis, G. and Elias-Rosa, N. and Firth, R. E. and Flewelling, H. and Flors, A. and Franckowiak, A. and Frohmaier, C. and Galbany, L. and Gonzalez-Gaitan, S. and Greiner, J. and Gromadzki, M. and Guelbenzu, A. Nicuesa and Gutierrez, C. P. and Hamanowicz, A. and Hanlon, L. and Harmanen, J. and Heintz, K. E. and Heinze, A. and Hernandez, M.-S. and Hodgkin, S. T. and Hook, I. M. and Izzo, L. and James, P. A. and Jonker, P. G. and Kerzendorf, W. E. and Klose, S. and Kostrzewa-Rutkowska, Z. and Kowalski, M. and Kromer, M. and Kuncarayakti, H. and Lawrence, A. and Lowe, T. B. and Magnier, E. A. and Manulis, I. and Martin-Carrillo, A. and Mattila, S. and McBrien, O. and Muller, A. and Nordin, J. and O'Neill, D. and Onori, F. and Palmerio, J. T. and Pastorello, A. and Patat, F. and Pignata, G. and Podsiadlowski, Ph and Pumo, M. L. and Prentice, S. J. and Rau, A. and Razza, A. and Rest, A. and Reynolds, T. and Roy, R. and Ruiter, A. J. and Rybicki, K. A. and Salmon, L. and Schady, P. and Schultz, A. S. B. and Schweyer, T. and Seitenzahl, I. R. and Smith, M. and Sollerman, J. and Stalder, B. and Stubbs, C. W. and Sullivan, M. and Szegedi, H. and Taddia, F. and Taubenberger, S. and Terreran, G. and van Soelen, B. and Vos, J. and Wainscoat, R. J. and Walton, N. A. and Waters, C. and Weiland, H. and Willman, M. and Wiseman, P. and Wright, D. E. and Wyrzykowski, L. and Yaron, O.},
	month = nov,
	year = {2017},
	note = {arXiv:1710.05841 [astro-ph]},
	pages = {75--79},
}

@article{garcia2020DESGW,
	title = {A {DESGW} {Search} for the {Electromagnetic} {Counterpart} to the {LIGO}/{Virgo} {Gravitational}-wave {Binary} {Neutron} {Star} {Merger} {Candidate} {S190510g}},
	volume = {903},
	issn = {1538-4357},
	url = {https://iopscience.iop.org/article/10.3847/1538-4357/abb823},
	doi = {10.3847/1538-4357/abb823},
	language = {en},
	number = {1},
	urldate = {2023-12-28},
	journal = {The Astrophysical Journal},
	author = {Garcia, A. and Morgan, R. and Herner, K. and Palmese, A. and Soares-Santos, M. and Annis, J. and Brout, D. and Vivas, A. K. and Drlica-Wagner, A. and Santana-Silva, L. and Tucker, D. L. and Allam, S. and Wiesner, M. and García-Bellido, J. and Gill, M. S. S. and Sako, M. and Kessler, R. and Davis, T. M. and Scolnic, D. and Casares, J. and Chen, H. and Conselice, C. and Cooke, J. and Doctor, Z. and Foley, R. J. and Horvath, J. and Howell, D. A. and Kilpatrick, C. D. and Lidman, C. and E., F. Olivares and Paz-Chinchón, F. and Pineda-G., J. and Quirola-Vásquez, J. and Rest, A. and Sherman, N. and Abbott, T. M. C. and Aguena, M. and Avila, S. and Bertin, E. and Bhargava, S. and Brooks, D. and Burke, D. L. and Rosell, A. Carnero and Kind, M. Carrasco and Carretero, J. and Costanzi, M. and Da Costa, L. N. and Desai, S. and Diehl, H. T. and Dietrich, J. P. and Doel, P. and Everett, S. and Flaugher, B. and Fosalba, P. and Friedel, D. and Frieman, J. and Gaztanaga, E. and Gerdes, D. W. and Gruen, D. and Gruendl, R. A. and Gschwend, J. and Gutierrez, G. and Hinton, S. R. and Hollowood, D. L. and Honscheid, K. and James, D. J. and Kuehn, K. and Kuropatkin, N. and Lahav, O. and Lima, M. and Maia, M. A. G. and March, M. and Marshall, J. L. and Menanteau, F. and Miquel, R. and Ogando, R. L. C. and Plazas, A. A. and Romer, A. K. and Roodman, A. and Sanchez, E. and Scarpine, V. and Schubnell, M. and Serrano, S. and Sevilla-Noarbe, I. and Smith, M. and Suchyta, E. and Swanson, M. E. C. and Tarle, G. and Thomas, D. and Varga, T. N. and Walker, A. R. and Weller, J. and {DES Collaboration}},
	month = nov,
	year = {2020},
	pages = {75},
}

@misc{dyer2020Telescope,
	title = {A telescope control and scheduling system for the {Gravitational}-wave {Optical} {Transient} {Observer}},
	url = {http://arxiv.org/abs/2003.06317},
	urldate = {2023-12-28},
	publisher = {arXiv},
	author = {Dyer, Martin J.},
	month = mar,
	year = {2020},
	note = {arXiv:2003.06317 [astro-ph]},
}

@article{abbott2017Gravitationalwave,
	title = {A gravitational-wave standard siren measurement of the {Hubble} constant},
	volume = {551},
	copyright = {2017 Macmillan Publishers Limited, part of Springer Nature. All rights reserved.},
	issn = {1476-4687},
	url = {https://www.nature.com/articles/nature24471},
	doi = {10.1038/nature24471},
	language = {en},
	number = {7678},
	urldate = {2023-10-17},
	journal = {Nature},
	publisher = {Nature Publishing Group},
	author = {Abbott, B. P. and Abbott, R. and Abbott, T. D. and Acernese, F. and Ackley, K. and Adams, C. and Adams, T. and Addesso, P. and Adhikari, R. X. and Adya, V. B. and Affeldt, C. and Afrough, M. and Agarwal, B. and Agathos, M. and Agatsuma, K. and Aggarwal, N. and Aguiar, O. D. and Aiello, L. and Ain, A. and Ajith, P. and Allen, B. and Allen, G. and Allocca, A. and Altin, P. A. and Amato, A. and Ananyeva, A. and Anderson, S. B. and Anderson, W. G. and Angelova, S. V. and Antier, S. and Appert, S. and Arai, K. and Araya, M. C. and Areeda, J. S. and Arnaud, N. and Arun, K. G. and Ascenzi, S. and Ashton, G. and Ast, M. and Aston, S. M. and Astone, P. and Atallah, D. V. and Aufmuth, P. and Aulbert, C. and AultONeal, K. and Austin, C. and Avila-Alvarez, A. and Babak, S. and Bacon, P. and Bader, M. K. M. and Bae, S. and Baker, P. T. and Baldaccini, F. and Ballardin, G. and Ballmer, S. W. and Banagiri, S. and Barayoga, J. C. and Barclay, S. E. and Barish, B. C. and Barker, D. and Barkett, K. and Barone, F. and Barr, B. and Barsotti, L. and Barsuglia, M. and Barta, D. and Bartlett, J. and Bartos, I. and Bassiri, R. and Basti, A. and Batch, J. C. and Bawaj, M. and Bayley, J. C. and Bazzan, M. and Bécsy, B. and Beer, C. and Bejger, M. and Belahcene, I. and Bell, A. S. and Berger, B. K. and Bergmann, G. and Bero, J. J. and Berry, C. P. L. and Bersanetti, D. and Bertolini, A. and Betzwieser, J. and Bhagwat, S. and Bhandare, R. and Bilenko, I. A. and Billingsley, G. and Billman, C. R. and Birch, J. and Birney, R. and Birnholtz, O. and Biscans, S. and Biscoveanu, S. and Bisht, A. and Bitossi, M. and Biwer, C. and Bizouard, M. A. and Blackburn, J. K. and Blackman, J. and Blair, C. D. and Blair, D. G. and Blair, R. M. and Bloemen, S. and Bock, O. and Bode, N. and Boer, M. and Bogaert, G. and Bohe, A. and Bondu, F. and Bonilla, E. and Bonnand, R. and Boom, B. A. and Bork, R. and Boschi, V. and Bose, S. and Bossie, K. and Bouffanais, Y. and Bozzi, A. and Bradaschia, C. and Brady, P. R. and Branchesi, M. and Brau, J. E. and Briant, T. and Brillet, A. and Brinkmann, M. and Brisson, V. and Brockill, P. and Broida, J. E. and Brooks, A. F. and Brown, D. A. and Brown, D. D. and Brunett, S. and Buchanan, C. C. and Buikema, A. and Bulik, T. and Bulten, H. J. and Buonanno, A. and Buskulic, D. and Buy, C. and Byer, R. L. and Cabero, M. and Cadonati, L. and Cagnoli, G. and Cahillane, C. and Bustillo, J. Calderón and Callister, T. A. and Calloni, E. and Camp, J. B. and Canepa, M. and Canizares, P. and Cannon, K. C. and Cao, H. and Cao, J. and Capano, C. D. and Capocasa, E. and Carbognani, F. and Caride, S. and Carney, M. F. and Diaz, J. Casanueva and Casentini, C. and Caudill, S. and Cavaglià, M. and Cavalier, F. and Cavalieri, R. and Cella, G. and Cepeda, C. B. and Cerdá-Durán, P. and Cerretani, G. and Cesarini, E. and Chamberlin, S. J. and Chan, M. and Chao, S. and Charlton, P. and Chase, E. and Chassande-Mottin, E. and Chatterjee, D. and Chatziioannou, K. and Cheeseboro, B. D. and Chen, H. Y. and Chen, X. and Chen, Y. and Cheng, H.-P. and Chia, H. and Chincarini, A. and Chiummo, A. and Chmiel, T. and Cho, H. S. and Cho, M. and Chow, J. H. and Christensen, N. and Chu, Q. and Chua, A. J. K. and Chua, S. and Chung, A. K. W. and Chung, S. and Ciani, G. and Ciolfi, R. and Cirelli, C. E. and Cirone, A. and Clara, F. and Clark, J. A. and Clearwater, P. and Cleva, F. and Cocchieri, C. and Coccia, E. and Cohadon, P.-F. and Cohen, D. and Colla, A. and Collette, C. G. and Cominsky, L. R. and Constancio, M. and Conti, L. and Cooper, S. J. and Corban, P. and Corbitt, T. R. and Cordero-Carrión, I. and Corley, K. R. and Cornish, N. and Corsi, A. and Cortese, S. and Costa, C. A. and Coughlin, M. W. and Coughlin, S. B. and Coulon, J.-P. and Countryman, S. T. and Couvares, P. and Covas, P. B. and Cowan, E. E. and Coward, D. M. and Cowart, M. J. and Coyne, D. C. and Coyne, R. and Creighton, J. D. E. and Creighton, T. D. and Cripe, J. and Crowder, S. G. and Cullen, T. J. and Cumming, A. and Cunningham, L. and Cuoco, E. and Dal Canton, T. and Dálya, G. and Danilishin, S. L. and D’Antonio, S. and Danzmann, K. and Dasgupta, A. and Da Silva Costa, C. F. and Datrier, L. E. H. and Dattilo, V. and Dave, I. and Davier, M. and Davis, D. and Daw, E. J. and Day, B. and De, S. and DeBra, D. and Degallaix, J. and De Laurentis, M. and Deléglise, S. and Del Pozzo, W. and Demos, N. and Denker, T. and Dent, T. and De Pietri, R. and Dergachev, V. and De Rosa, R. and DeRosa, R. T. and De Rossi, C. and DeSalvo, R. and de Varona, O. and Devenson, J. and Dhurandhar, S. and Díaz, M. C. and Di Fiore, L. and Di Giovanni, M. and Di Girolamo, T. and Di Lieto, A. and Di Pace, S. and Di Palma, I. and Di Renzo, F. and Doctor, Z. and Dolique, V. and Donovan, F. and Dooley, K. L. and Doravari, S. and Dorrington, I. and Douglas, R. and Dovale álvarez, M. and Downes, T. P. and Drago, M. and Dreissigacker, C. and Driggers, J. C. and Du, Z. and Ducrot, M. and Dupej, P. and Dwyer, S. E. and {The LIGO Scientific Collaboration and The Virgo Collaboration}},
	month = nov,
	year = {2017},
	note = {Number: 7678},
	pages = {85--88},
}

@article{goldstein2017Ordinary,
	title = {An {Ordinary} {Short} {Gamma}-{Ray} {Burst} with {Extraordinary} {Implications}: {Fermi}-{GBM} {Detection} of {GRB} {170817A}},
	volume = {848},
	issn = {2041-8205},
	shorttitle = {An {Ordinary} {Short} {Gamma}-{Ray} {Burst} with {Extraordinary} {Implications}},
	url = {https://dx.doi.org/10.3847/2041-8213/aa8f41},
	doi = {10.3847/2041-8213/aa8f41},
	language = {en},
	number = {2},
	urldate = {2023-09-30},
	journal = {The Astrophysical Journal Letters},
	publisher = {The American Astronomical Society},
	author = {Goldstein, A. and Veres, P. and Burns, E. and Briggs, M. S. and Hamburg, R. and Kocevski, D. and Wilson-Hodge, C. A. and Preece, R. D. and Poolakkil, S. and Roberts, O. J. and Hui, C. M. and Connaughton, V. and Racusin, J. and Kienlin, A. von and Canton, T. Dal and Christensen, N. and Littenberg, T. and Siellez, K. and Blackburn, L. and Broida, J. and Bissaldi, E. and Cleveland, W. H. and Gibby, M. H. and Giles, M. M. and Kippen, R. M. and McBreen, S. and McEnery, J. and Meegan, C. A. and Paciesas, W. S. and Stanbro, M.},
	month = oct,
	year = {2017},
	pages = {L14},
}

@article{rossi2020Comparison,
	title = {A comparison between short {GRB} afterglows and kilonova {AT2017gfo}: shedding light on kilonovae properties},
	volume = {493},
	issn = {0035-8711, 1365-2966},
	shorttitle = {A comparison between short {GRB} afterglows and kilonova {AT2017gfo}},
	url = {https://academic.oup.com/mnras/article/493/3/3379/5740732},
	doi = {10.1093/mnras/staa479},
	language = {en},
	number = {3},
	urldate = {2023-08-04},
	journal = {Monthly Notices of the Royal Astronomical Society},
	author = {Rossi, A and Stratta, G and Maiorano, E and Spighi, D and Masetti, N and Palazzi, E and Gardini, A and Melandri, A and Nicastro, L and Pian, E and Branchesi, M and Dadina, M and Testa, V and Brocato, E and Benetti, S and Ciolfi, R and Covino, S and D’Elia, V and Grado, A and Izzo, L and Perego, A and Piranomonte, S and Salvaterra, R and Selsing, J and Tomasella, L and Yang, S and Vergani, D and Amati, L and Stephen, J B},
	month = apr,
	year = {2020},
	pages = {3379--3397},
}

@article{rastinejad2022Kilonova,
	title = {A {Kilonova} {Following} a {Long}-{Duration} {Gamma}-{Ray} {Burst} at 350 {Mpc}},
	volume = {612},
	issn = {0028-0836, 1476-4687},
	url = {http://arxiv.org/abs/2204.10864},
	doi = {10.1038/s41586-022-05390-w},
	number = {7939},
	urldate = {2023-03-23},
	journal = {Nature},
	author = {Rastinejad, J. C. and Gompertz, B. P. and Levan, A. J. and Fong, W. and Nicholl, M. and Lamb, G. P. and Malesani, D. B. and Nugent, A. E. and Oates, S. R. and Tanvir, N. R. and Postigo, A. de Ugarte and Kilpatrick, C. D. and Moore, C. J. and Metzger, B. D. and Ravasio, M. E. and Rossi, A. and Schroeder, G. and Jencson, J. and Sand, D. J. and Smith, N. and Fernández, J. F. Agüí and Berger, E. and Blanchard, P. K. and Chornock, R. and Cobb, B. E. and De Pasquale, M. and Fynbo, J. P. U. and Izzo, L. and Kann, D. A. and Laskar, T. and Marini, E. and Paterson, K. and Escorial, A. Rouco and Sears, H. M. and Thöne, C. C.},
	month = dec,
	year = {2022},
	note = {arXiv:2204.10864 [astro-ph]},
	pages = {223--227},
}

@article{frostig2022Infrared,
	title = {An {Infrared} {Search} for {Kilonovae} with the {WINTER} {Telescope}. {I}. {Binary} {Neutron} {Star} {Mergers}},
	volume = {926},
	issn = {0004-637X, 1538-4357},
	url = {https://iopscience.iop.org/article/10.3847/1538-4357/ac4508},
	doi = {10.3847/1538-4357/ac4508},
	language = {en},
	number = {2},
	urldate = {2023-02-02},
	journal = {The Astrophysical Journal},
	author = {Frostig, Danielle and Biscoveanu, Sylvia and Mo, Geoffrey and Karambelkar, Viraj and Dal Canton, Tito and Chen, Hsin-Yu and Kasliwal, Mansi and Katsavounidis, Erik and Lourie, Nathan P. and Simcoe, Robert A. and Vitale, Salvatore},
	month = feb,
	year = {2022},
	pages = {152},
}

@article{berger2013RProcess,
	title = {An r-{Process} {Kilonova} {Associated} with the {Short}-{Hard} {GRB} {130603B}},
	volume = {774},
	issn = {2041-8205, 2041-8213},
	url = {http://arxiv.org/abs/1306.3960},
	doi = {10.1088/2041-8205/774/2/L23},
	number = {2},
	urldate = {2024-03-27},
	journal = {The Astrophysical Journal},
	author = {Berger, E. and Fong, W. and Chornock, R.},
	month = aug,
	year = {2013},
	note = {arXiv:1306.3960 [astro-ph]},
	pages = {L23},
}

@article{li1998Transient,
	title = {Transient {Events} from {Neutron} {Star} {Mergers}},
	volume = {507},
	issn = {0004-637X},
	url = {https://iopscience.iop.org/article/10.1086/311680/meta},
	doi = {10.1086/311680},
	language = {en},
	number = {1},
	urldate = {2024-03-27},
	journal = {The Astrophysical Journal},
	publisher = {IOP Publishing},
	author = {Li, Li-Xin and Paczyński, Bohdan},
	month = sep,
	year = {1998},
	pages = {L59},
}

@article{pian2017Spectroscopic,
	title = {Spectroscopic identification of r-process nucleosynthesis in a double neutron star merger},
	volume = {551},
	issn = {0028-0836, 1476-4687},
	url = {http://arxiv.org/abs/1710.05858},
	doi = {10.1038/nature24298},
	number = {7678},
	urldate = {2024-03-27},
	journal = {Nature},
	author = {Pian, E. and D'Avanzo, P. and Benetti, S. and Branchesi, M. and Brocato, E. and Campana, S. and Cappellaro, E. and Covino, S. and D'Elia, V. and Fynbo, J. P. U. and Getman, F. and Ghirlanda, G. and Ghisellini, G. and Grado, A. and Greco, G. and Hjorth, J. and Kouveliotou, C. and Levan, A. and Limatola, L. and Malesani, D. and Mazzali, P. A. and Melandri, A. and Moller, P. and Nicastro, L. and Palazzi, E. and Piranomonte, S. and Rossi, A. and Salafia, O. S. and Selsing, J. and Stratta, G. and Tanaka, M. and Tanvir, N. R. and Tomasella, L. and Watson, D. and Yang, S. and Amati, L. and Antonelli, L. A. and Ascenzi, S. and Bernardini, M. G. and Boer, M. and Bufano, F. and Bulgarelli, A. and Capaccioli, M. and Casella, P. G. and Castro-Tirado, A. J. and Chassande-Mottin, E. and Ciolfi, R. and Copperwheat, C. M. and Dadina, M. and De Cesare, G. and Di Paola, A. and Fan, Y. Z. and Gendre, B. and Giuffrida, G. and Giunta, A. and Hunt, L. K. and Israel, G. and Jin, Z.-P. and Kasliwal, M. and Klose, S. and Lisi, M. and Longo, F. and Maiorano, E. and Mapelli, M. and Masetti and Nava, L. and Patricelli, B. and Perley, D. and Pescalli, A. and Piran, T. and Possenti, A. and Pulone, L. and Razzano, M. and Salvaterra, R. and Schipani, P. and Spera, M. and Stamerra, A. and Stella, L. and Tagliaferri, G. and Testa, V. and Troja, E. and Turatto, M. and Vergani, S. D. and Vergani, D.},
	month = nov,
	year = {2017},
	note = {arXiv:1710.05858 [astro-ph]},
	pages = {67--70},
}

@article{metzger2020Kilonovae,
	title = {Kilonovae},
	volume = {23},
	issn = {14338351},
	doi = {10.1007/s41114-019-0024-0},
	number = {1},
	journal = {Living Reviews in Relativity},
	author = {Metzger, Brian D.},
	year = {2020},
	note = {arXiv: 1910.01617},
}

@article{kasliwal2017Illuminating,
	title = {Illuminating gravitational waves: {A} concordant picture of photons from a neutron star merger},
	volume = {358},
	shorttitle = {Illuminating gravitational waves},
	url = {https://www.science.org/doi/10.1126/science.aap9455},
	doi = {10.1126/science.aap9455},
	number = {6370},
	urldate = {2024-03-27},
	journal = {Science},
	publisher = {American Association for the Advancement of Science},
	author = {Kasliwal, M. M. and Nakar, E. and Singer, L. P. and Kaplan, D. L. and Cook, D. O. and Van Sistine, A. and Lau, R. M. and Fremling, C. and Gottlieb, O. and Jencson, J. E. and Adams, S. M. and Feindt, U. and Hotokezaka, K. and Ghosh, S. and Perley, D. A. and Yu, P.-C. and Piran, T. and Allison, J. R. and Anupama, G. C. and Balasubramanian, A. and Bannister, K. W. and Bally, J. and Barnes, J. and Barway, S. and Bellm, E. and Bhalerao, V. and Bhattacharya, D. and Blagorodnova, N. and Bloom, J. S. and Brady, P. R. and Cannella, C. and Chatterjee, D. and Cenko, S. B. and Cobb, B. E. and Copperwheat, C. and Corsi, A. and De, K. and Dobie, D. and Emery, S. W. K. and Evans, P. A. and Fox, O. D. and Frail, D. A. and Frohmaier, C. and Goobar, A. and Hallinan, G. and Harrison, F. and Helou, G. and Hinderer, T. and Ho, A. Y. Q. and Horesh, A. and Ip, W.-H. and Itoh, R. and Kasen, D. and Kim, H. and Kuin, N. P. M. and Kupfer, T. and Lynch, C. and Madsen, K. and Mazzali, P. A. and Miller, A. A. and Mooley, K. and Murphy, T. and Ngeow, C.-C. and Nichols, D. and Nissanke, S. and Nugent, P. and Ofek, E. O. and Qi, H. and Quimby, R. M. and Rosswog, S. and Rusu, F. and Sadler, E. M. and Schmidt, P. and Sollerman, J. and Steele, I. and Williamson, A. R. and Xu, Y. and Yan, L. and Yatsu, Y. and Zhang, C. and Zhao, W.},
	month = dec,
	year = {2017},
	pages = {1559--1565},
}

@article{barnes2013EFFECT,
	title = {{EFFECT} {OF} {A} {HIGH} {OPACITY} {ON} {THE} {LIGHT} {CURVES} {OF} {RADIOACTIVELY} {POWERED} {TRANSIENTS} {FROM} {COMPACT} {OBJECT} {MERGERS}},
	volume = {775},
	issn = {0004-637X},
	url = {https://dx.doi.org/10.1088/0004-637X/775/1/18},
	doi = {10.1088/0004-637X/775/1/18},
	language = {en},
	number = {1},
	urldate = {2024-03-27},
	journal = {The Astrophysical Journal},
	publisher = {The American Astronomical Society},
	author = {Barnes, Jennifer and Kasen, Daniel},
	month = aug,
	year = {2013},
	pages = {18},
}

% Alternatively, you could enter them by hand, like this:
%\begin{thebibliography}{99}
%\end{thebibliography}

%%%%%%%%%%%%%%%%%%%%%%%%%%%%%%%%%%%%%%%%%%%%%%%%%%

%%%%%%%%%%%%%%%%% APPENDICES %%%%%%%%%%%%%%%%%%%%%

% \appendix

% \section{Some extra material}

%%%%%%%%%%%%%%%%%%%%%%%%%%%%%%%%%%%%%%%%%%%%%%%%%%

% Don't change these lines
\bsp% typesetting comment
\label{lastpage}
\end{document}